\documentclass[12pt]{article}
\usepackage{amssymb}
\usepackage{amsmath}
\usepackage{amsthm}
\usepackage{color}
\usepackage{comment}
\usepackage{geometry}		
\usepackage{graphicx}
\usepackage{float}
\usepackage{booktabs,longtable}
\usepackage[english]{babel}
\usepackage{multirow}
\usepackage{tabularx,booktabs}
\usepackage{caption}
\usepackage{subcaption}
\usepackage{hyperref}
\hypersetup{
	colorlinks=true,
	linkcolor=blue,
	urlcolor=blue,
	citecolor=blue
}

\usepackage{mathtools}
\usepackage{tensor}

\theoremstyle{definition}

\theoremstyle{remark}

\usepackage{authblk} 
\usepackage{orcidlink} 
\usepackage{hyperref}
\hypersetup{colorlinks=true, linkcolor=blue, urlcolor=blue}

\usepackage{multicol}
\usepackage{changepage}
\let\originalleft\left
\let\originalright\right
\renewcommand{\left}{\mathopen{}\mathclose\bgroup\originalleft}
\renewcommand{\right}{\aftergroup\egroup\originalright}

\usepackage{tabularx}
\usepackage{caption}
\usepackage{subcaption}

\usepackage{tikz}
\usetikzlibrary{arrows.meta, positioning,backgrounds}

\usepackage{placeins}
\usepackage{setspace}
\usepackage{pdflscape}

\title{Spatiotemporal dynamics and ecological risk factors of highly pathogenic avian influenza A(H5N1) in Canadian wildlife: A One Health surveillance analysis}

\author[1]{H.O.~Fatoyinbo\,\orcidlink{0000-0002-6036-2957}\thanks{Corresponding author: \href{mailto:hammed.fatoyinbo@aut.ac.nz}{hammed.fatoyinbo@aut.ac.nz}}}
\author[1]{Hoyeon.~Jeong}

\affil[1]{Department of Mathematical Sciences, Auckland University of Technology, Auckland 1010, New Zealand}
\begin{document}
	\maketitle
	
	
	\begin{abstract}
	Highly pathogenic avian influenza A(H5N1) has expanded geographically and ecologically, affecting wild birds, mammalian wildlife, domestic animals, and humans. Wildlife surveillance provides critical early warning for One Health preparedness, yet national-scale analyses integrating host ecology, spatial patterns, seasonality, viral lineage, and risk factors remain limited. This study analysed Canadian wildlife HPAI A(H5N1) surveillance records from 2022 to 2026 to characterise spatiotemporal dynamics and identify factors associated with detection counts. A retrospective analysis of 2,657 detections across 13 provinces and territories was conducted using descriptive epidemiology, spatial clustering methods, and Negative Binomial mixed models. Detections were predominantly avian (93.3\%), with waterfowl and raptors as the major host groups, while mammals accounted for a smaller but epidemiologically important proportion (6.7\%). Detection burden was highest in 2022, with increased activity in autumn and spring. Ontario, Alberta, and British Columbia were identified as major hotspots, with evidence of local clustering in parts of the Prairie region. Reassortant Eurasian–North American lineages dominated detections (80.8\%) and were strongly associated with higher detection counts. Modelling results identified year, season, and lineage as key predictors. These findings support risk-based One Health surveillance prioritising high-burden regions, migration-associated periods, key avian host groups, reassortant viral lineages, and continued monitoring of mammalian wildlife.
	\end{abstract}

	
\section{Introduction}

Highly pathogenic avian influenza A(H5N1) has become an expanding One Health challenge because its transmission dynamics involve wild birds, domestic poultry, mammalian wildlife, humans, and shared environments. The One Health approach recognises the interdependence of human, animal, and ecosystem health and emphasises cross-sector collaboration for prevention, detection, preparedness, and response \cite{WHO2024OneHealth,FAO2024OneHealth}. In avian influenza, this framework is especially relevant because viral maintenance and spillover occur across ecological interfaces linking migratory birds, domestic animals, contaminated environments, and occupationally exposed human populations \cite{FAO2025AvianInfluenza,Mena2025H5N1USReview}.

Wild birds are central to avian influenza ecology. Waterbirds are recognised as important reservoirs, while viral persistence and transmission are shaped by host species, migration, environmental conditions, and contact with domestic birds \cite{Causey2008AIVEcology,Spackman2009AIVEcologyPoultry,Vandegrift2010AIVChangingWorld}. Long-distance dispersal of avian influenza viruses has also been linked to migratory bird networks and intercontinental movement pathways \cite{Kilpatrick2006GlobalSpreadH5N1,Tian2015H5N1MigrationNetworks}. The recent HPAI H5 clade 2.3.4.4b panzootic has altered the geographic range and host breadth of HPAI outbreaks, causing extensive mortality in wild birds, outbreaks in poultry, and increasing spillovers into mammals \cite{Couty2026WildBirdsPanzootic,Mena2025H5N1USReview}.

The arrival of HPAI A(H5N1) clade 2.3.4.4b in North America was closely linked to wild-bird movement. In December 2021, H5N1 HPAI viruses were detected in poultry and a free-living gull in St. John's, Newfoundland and Labrador, with phylogenetic evidence consistent with transatlantic movement from northwestern Europe \cite{Caliendo2022TransatlanticH5N1}. Genomic analyses in the United States subsequently identified multiple wild-bird introductions and extensive reassortment with North American avian influenza viruses \cite{Youk2023H5N1ReassortmentUS}. In Canada, national surveillance during the first year after incursion documented HPAIV detections across all provinces and territories, diverse taxonomic groups, and both Eurasian-origin and reassortant viral lineages \cite{Giacinti2024CanadaWildBirds}.

The Canadian outbreak also had major conservation implications, especially for seabirds in eastern Canada. Large-scale mortality assessments estimated more than 40,000 HPAI-attributable wild-bird deaths in eastern Canada during spring and summer 2022, with seabirds and sea ducks accounting for most estimated mortalities \cite{AveryGomm2024EasternCanadaMortalities}. Live-bird surveillance in Atlantic Canada further showed evidence of active infection and elevated H5 antibody seroprevalence in seabird populations after the incursion, highlighting the value of combining carcass-based surveillance with live-bird and serological monitoring \cite{Rahman2026LiveBirdSeabirdH5N1}. Seabird impacts have also been documented internationally, including severe demographic consequences in Northern Gannets during the current H5N1 panzootic \cite{Lane2024GannetsH5N1}. International comparisons also show that avian influenza ecology is context-dependent; for example, long-term surveillance in New Zealand found low-pathogenic H5 and H7 viruses mainly in resident mallards, with limited evidence of influenza viruses in major migratory bird populations visiting New Zealand \cite{Stanislawek2024AIVNewZealand}.

Risk-based HPAI surveillance increasingly relies on integrating wildlife, environmental, poultry, and movement data. Adaptive and predictive modelling studies have shown that waterfowl abundance, avian influenza prevalence, poultry density, environmental conditions, and climate-related variables can help identify spatiotemporal outbreak risk \cite{Prosser2024AdaptiveHPAI,Opata2025HPAIEuropePredictive,Zou2026ExtremeWeatherH5N1}. Environmental drivers such as temperature, hydrology, habitat, and land use may also influence viral persistence, host movement, and outbreak risk \cite{Fang2008EnvironmentalH5N1China,Morin2018AIVClimateLens}. These findings support the use of integrated spatiotemporal and ecological modelling to guide surveillance priorities.

Despite growing evidence on HPAI A(H5N1), there remains a need for national-scale analyses that integrate Canadian wildlife surveillance data across multiple years, provinces, host categories, seasons, migration periods, and viral lineages. This study analysed Canadian wildlife HPAI A(H5N1) surveillance records from 2022 to early 2026. The objectives were to: (i) describe temporal and seasonal patterns of wildlife detections; (ii) identify province-level spatial hotspots and local clustering; (iii) characterise host-category and viral lineage patterns; and (iv) model ecological and temporal risk factors associated with detection counts using Negative Binomial generalised linear mixed models. The study was framed as a One Health surveillance analysis with implications for preventive monitoring, preparedness, and targeted risk communication.

\section{Methods}

\subsection{Study Design}

This retrospective observational study analysed Canadian wildlife HPAI A(H5N1) surveillance detections to describe temporal, spatial, host-associated, and virological patterns and to identify predictors of detection counts within a One Health surveillance framework.


\subsection{Data Source}

Wildlife HPAI surveillance data for Canada were obtained from the public Highly Pathogenic Avian Influenza wildlife dashboard hosted through the Canadian Food Inspection Agency/ArcGIS platform and linked by the Canadian Wildlife Health Cooperative \cite{CFIAHPAIDashboards2025,CWHCAvianInfluenzaDashboard2025,ArcGISWildPositivesDashboard2026}. The dashboard provides confirmed HPAI detections in Canadian wildlife and can be searched by species or species group, province, collection date, and animal status at sampling. The original dashboard URL used for data extraction was: \url{https://cfia-ncr.maps.arcgis.com/apps/dashboards/89c779e98cdf492c899df23e1c38fdbc}. The dataset contained records of HPAI detections in free-ranging wildlife, including province, collection date, result authorisation date, animal class, bird group, common name, scientific name, infection status, viral strain, and lineage. Each record represented a wildlife detection event. Individual records were used for descriptive summaries, while aggregated detection counts were used for spatiotemporal analysis and risk-factor modelling.

\subsection{Data Cleaning and Preparation}


Raw data were imported into R, column names were standardised, dates were converted to date format, and categorical variables were harmonised. Missing categorical values were recoded as ``Unknown'' where appropriate. Exact duplicate rows were removed, followed by removal of likely epidemiological duplicates using province, common name, scientific name, collection date, status, strain, and lineage. Records from 2021 were excluded because detection counts were too small for stable annual comparisons.

\subsection{Derived Variables}




Temporal variables included year, month, ISO week, season, and migration period. Seasons were defined as winter, December--February; spring, March--May; summer, June--August; and autumn, September--November. Spring and fall migration periods were defined as March--May and September--November, respectively.

Province was the primary spatial unit. Province-level detection counts were joined to Canadian province and territory boundaries for mapping and spatial autocorrelation analysis. Host variables included animal class, common name, scientific name, bird group, host type, and bird category. Avian records were classified as birds and mammalian records as mammals. Mortality-associated detections included records coded as ``Dead'' or ``Euthanized''.

\subsection{Descriptive and Spatiotemporal Analysis}

Descriptive analyses summarised detections across temporal, spatial, host, and virological variables. Temporal patterns were visualised using annual, monthly, and weekly epidemic curves.

Province-level hotspot classification was performed using quartiles of total detection counts. Provinces and territories were classified into low, moderate, high, and very high detection categories. Spatial patterns were mapped using Canadian province and territory boundary data.

Spatial autocorrelation was assessed using Global Moran's I and Local Indicators of Spatial Association (LISA) to evaluate overall and local clustering of province-level detection counts \cite{Moran1950SpatialAutocorrelation,Anselin1995LISA}. Spatial weights were constructed using k-nearest-neighbour relationships based on province and territory geometry. Local Moran's I results were classified as high-high, low-low, high-low, low-high, or not significant.

\subsection{Risk-Factor Modelling}

Risk-factor modelling was conducted using generalised linear mixed models to identify predictors associated with HPAI detection counts. Because detection counts were overdispersed and clustered by province and host species, Negative Binomial generalised linear mixed models were used as the main modelling approach \cite{Brooks2017glmmTMB}. The selection of temporal, spatial, host, and virological predictors was informed by previous HPAI risk-factor studies showing that outbreak occurrence and detection intensity vary by season, geography, host category, poultry and wild-bird ecology, and environmental conditions \cite{Fatoyinbo2025HPAIAsia,Prosser2024AdaptiveHPAI,Opata2025HPAIEuropePredictive}.

The primary outcome was detection count per province--time--host--lineage stratum. Detection records were aggregated by province, year, season, migration period, host category, species, host type, and lineage. Rare species with fewer than five detections were collapsed into an ``Other Rare Species'' category. Rare lineages with fewer than ten detections were collapsed into an ``Other/Rare Lineage'' category.

The main season-based Negative Binomial model was specified as:
\begin{equation}
	Y_i \sim \text{NegBin}(\mu_i, \theta),
\end{equation}

\begin{equation}
	\log(\mu_i) =
	\beta_0 +
	\beta_1\text{Year}_i +
	\beta_2\text{Season}_i +
	\beta_3\text{BirdCategory}_i +
	\beta_4\text{Lineage}_i +
	\beta_5\text{HostType}_i +
	u_{\text{province}[i]} +
	u_{\text{species}[i]}.
\end{equation}

A migration-period sensitivity model was fitted by replacing season with migration period. Season and migration period were not included in the same main model because migration period was derived from month and strongly overlapped with seasonal classification. A Poisson GLMM was fitted as a baseline comparison model. A zero-inflated Negative Binomial model was also fitted to assess whether excess zeros were better accounted for using a zero-inflation component. Model fit was compared using Akaike Information Criterion. Residual diagnostics were conducted using simulation-based residuals in \texttt{DHARMa}, including tests for overdispersion and zero inflation \cite{Hartig2024DHARMa}. Fixed-effect estimates were exponentiated and reported as incidence rate ratios (IRRs) with 95\% confidence intervals.

\subsection{Software}

Analyses were performed in R using \texttt{tidyverse} for data processing and visualisation, \texttt{sf} and \texttt{spdep} for spatial analysis, \texttt{glmmTMB} for mixed-effects count models, and \texttt{DHARMa} for residual diagnostics.
\subsection{Ethical Considerations}

This study used secondary surveillance data and did not involve direct contact with animals or humans. No personally identifiable human information was analysed. The analysis was conducted at aggregated spatial, temporal, and host-group levels to support wildlife disease surveillance, One Health risk interpretation, and preventive public-health preparedness.

\section{Results}

\subsection{Dataset Overview}

After cleaning, deduplication, and exclusion of 2021 records, the final analytical dataset contained 2,657 HPAI A(H5N1) wildlife detections from 13 Canadian provinces and territories between 7 January 2022 and 25 March 2026. The dataset included 112 wildlife species grouped into eight host categories. Most detections were avian (n = 2,478; 93.3\%), while mammals accounted for 179 detections (6.7\%). Mortality-associated records accounted for 1,980 detections (74.5\%).

\subsection{Temporal and Seasonal Distribution of HPAI Detections}

Annual detection counts varied substantially across the study period. The highest number of detections occurred in 2022, with 1,199 records, representing 45.1\% of the final analytical dataset. Detection counts were lower in 2023 and 2024, with 423 and 363 detections, respectively, before increasing again in 2025 to 605 detections. A smaller number of detections was recorded in 2026, reflecting partial-year surveillance through March 2026.

\begin{figure}[ht]
	\centering
	\includegraphics[width=0.8\textwidth]{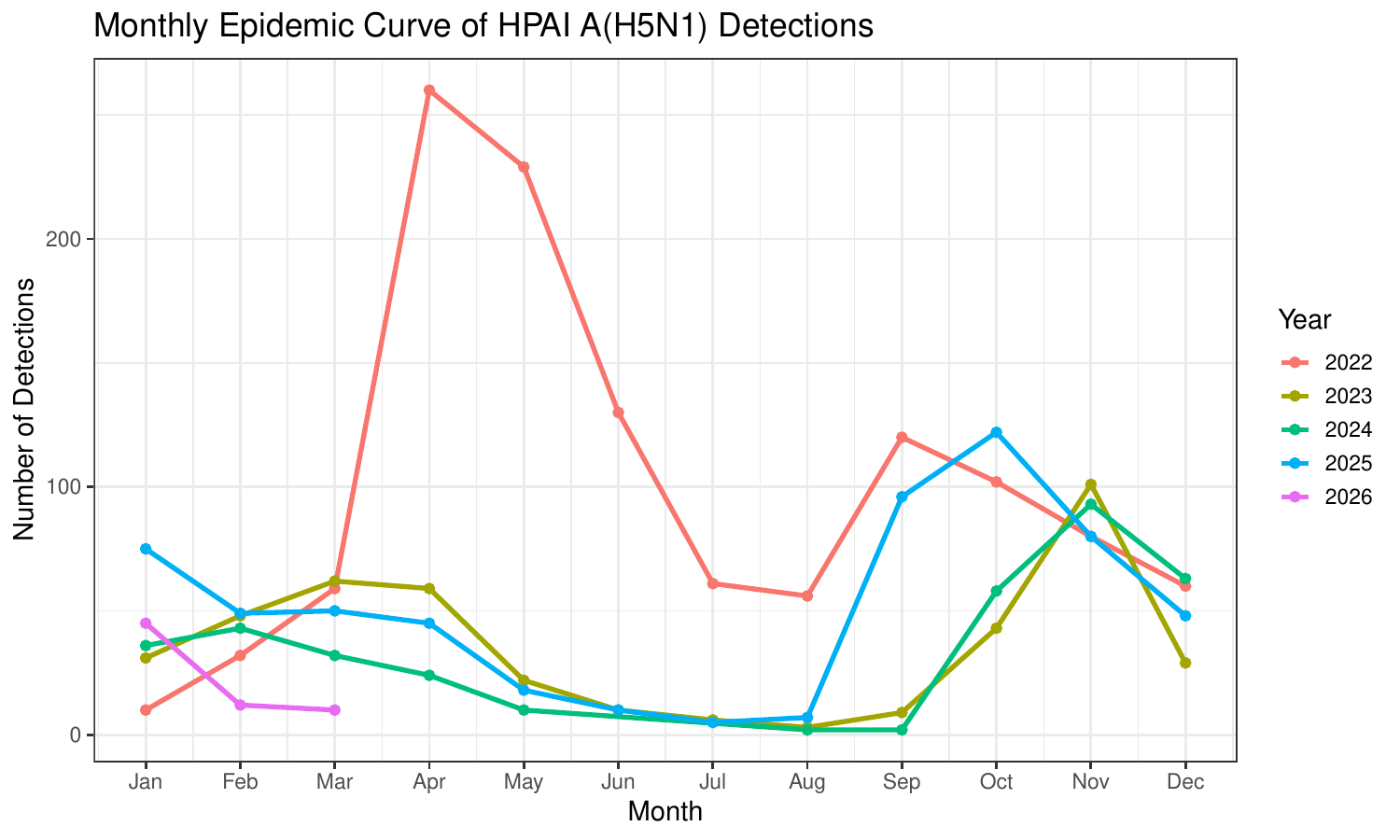}
	\caption{
		Monthly epidemic curve of HPAI A(H5N1) detections in Canadian wildlife, 2022--2026. Detection counts varied substantially by year and month, with a pronounced spring peak in 2022 and recurring detection activity in later years. The 2026 data represent partial-year surveillance through March 2026.
	}
	\label{fig:monthly_epidemic_curve}
\end{figure}

The monthly epidemic curve showed pronounced temporal heterogeneity, with a large spring peak in 2022 and recurring detection activity in later years (Figure~\ref{fig:monthly_epidemic_curve}). Weekly epidemic curves supported the presence of distinct epidemic waves, although monthly aggregation provided a clearer summary of the main temporal pattern. Additional annual, weekly, and threshold-based epidemic-wave plots are provided in Supplementary Figures S1--S3.

Seasonal analysis showed that detections were unevenly distributed across seasons. Autumn had the highest number of detections, with 906 records, followed closely by spring with 880 detections. Winter accounted for 581 detections, while summer had the lowest detection count, with 290 records. These represented 34.1\%, 33.1\%, 21.9\%, and 10.9\% of detections, respectively.

Migration-period classification showed a similar distribution. Fall migration accounted for 906 detections, spring migration accounted for 880 detections, and non-migration periods accounted for 871 detections. These represented 34.1\%, 33.1\%, and 32.8\% of detections, respectively. Although fall and spring migration periods were important contributors to detection activity, the substantial number of detections during non-migration periods indicates that HPAI A(H5N1) surveillance should not be restricted to migration seasons alone.

Detailed seasonal and migration-period summaries are provided in Supplementary Table S2. Province-level migration-period variation is shown in Supplementary Figure S9.

\subsection{Spatial Hotspots and Geographic Clustering}

\begin{figure}[ht]
	\centering
	\includegraphics[width=0.48\textwidth]{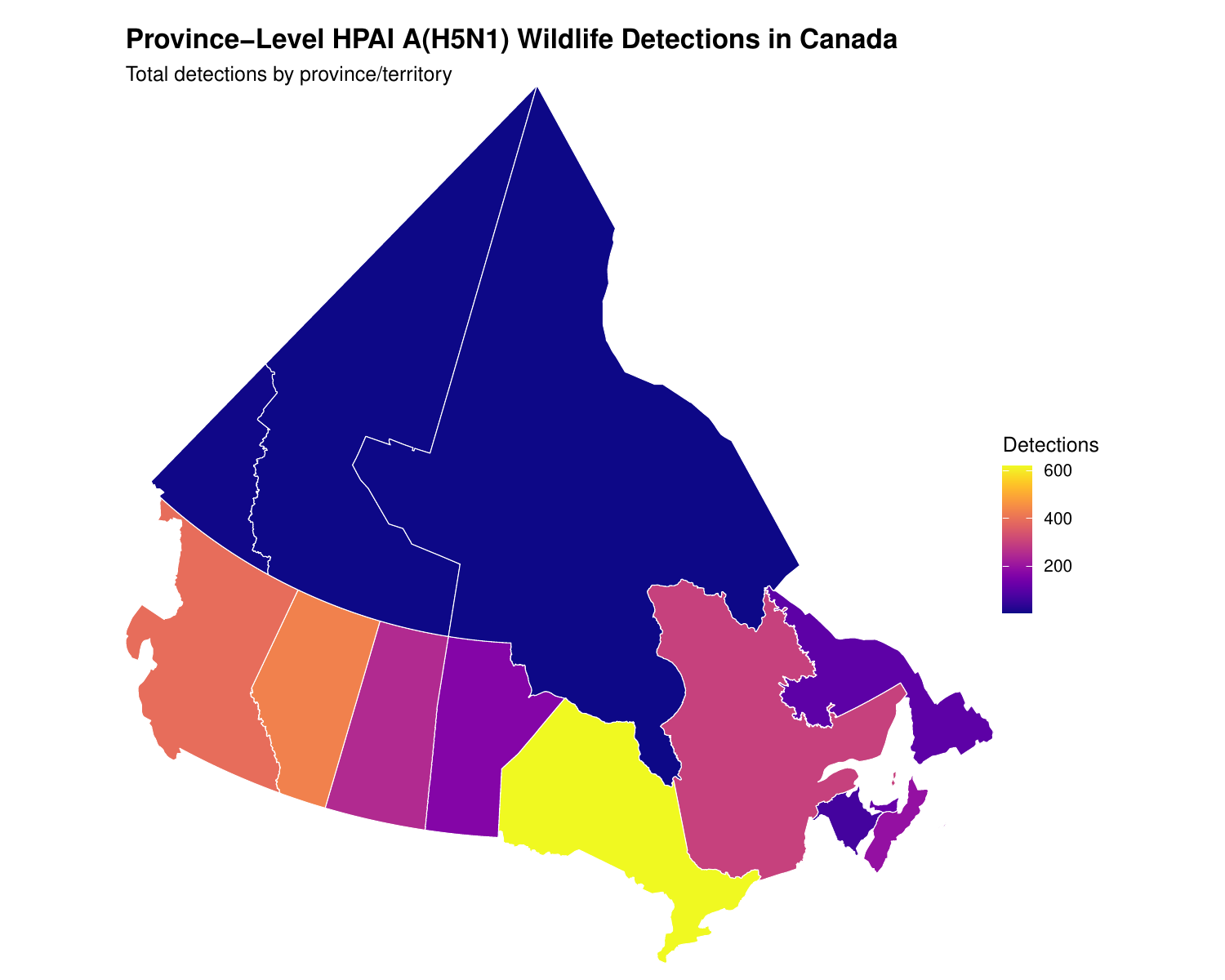}
	\includegraphics[width=0.48\textwidth]{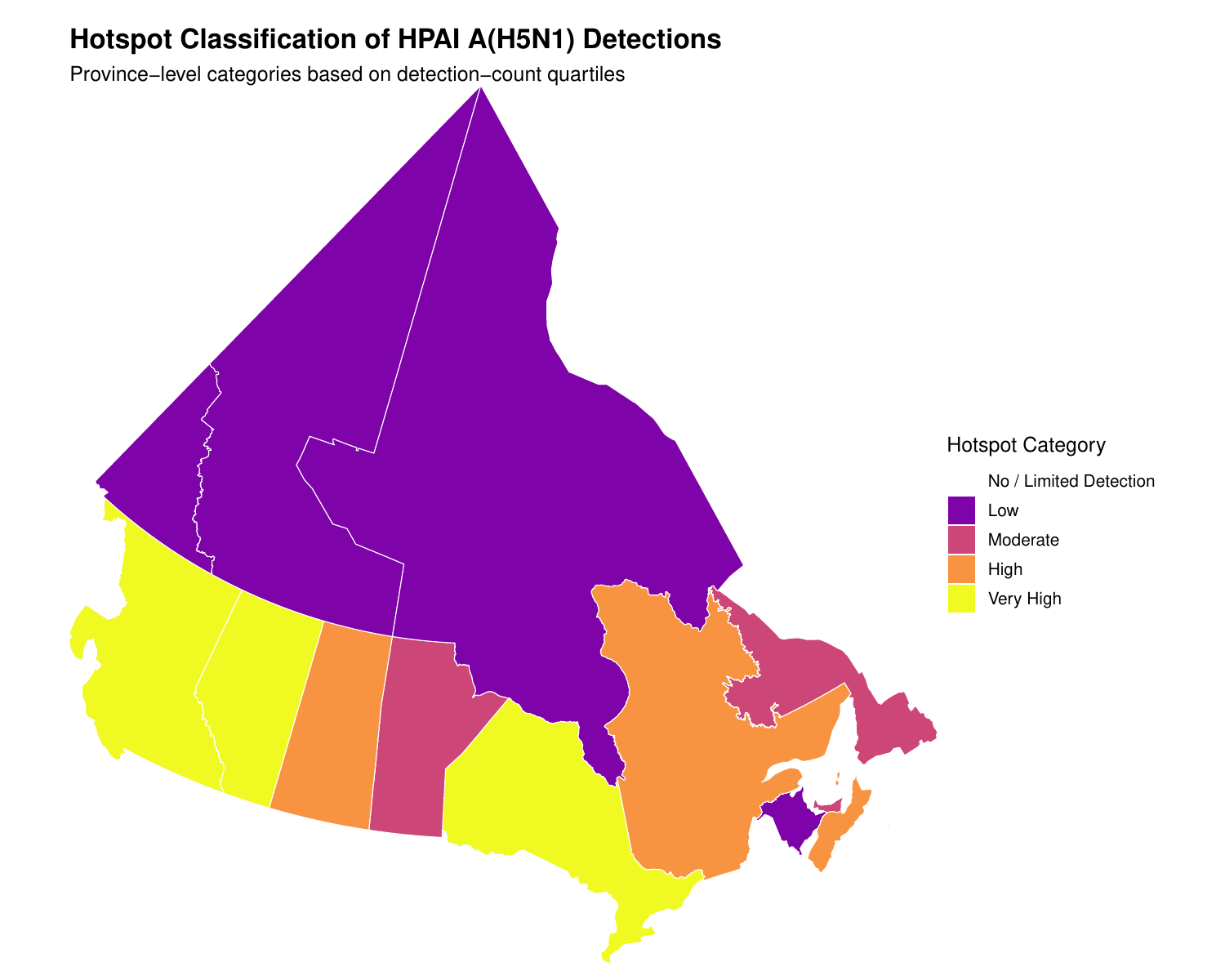}
	\caption{
		Spatial distribution and hotspot classification of HPAI A(H5N1) detections in Canadian wildlife. Left: province- and territory-level detection counts aggregated across the study period. Right: hotspot categories based on quartiles of total detection counts. The maps show marked geographic heterogeneity, with detections concentrated in a subset of provinces.
	}
	\label{fig:spatial_hotspots}
\end{figure}

Province-level mapping showed marked geographic heterogeneity in HPAI A(H5N1) wildlife detections across Canada (Figure~\ref{fig:spatial_hotspots}). Ontario recorded the highest number of detections (n = 618), followed by Alberta (n = 429) and British Columbia (n = 391). These three provinces were classified as very high detection hotspots. Quebec (n = 298), Saskatchewan (n = 249), and Nova Scotia (n = 195) were classified as high detection provinces. Manitoba, Prince Edward Island, and Newfoundland and Labrador were classified as moderate detection provinces, while New Brunswick, Yukon, Northwest Territories, and Nunavut were classified as low detection areas.

Global Moran's I did not provide evidence of statistically significant overall spatial autocorrelation in province-level detection counts (Moran's I = -0.0176, expected I = -0.0833, p = 0.3266). Local Moran's I identified one significant high-high cluster in Saskatchewan (Local Moran's I = 0.2697, z = 2.5109, p = 0.0120) and a significant low-high spatial outlier in Manitoba (Local Moran's I = -0.2495, z = -2.6800, p = 0.0074). All other provinces and territories were not statistically significant in the Local Moran's I analysis. Additional province-year, province-month, province-season, and region-season heatmaps are provided in Supplementary Figures S5--S8. The Local Moran's I cluster map is provided in Supplementary Figure S15, and significant Local Moran's I results are summarised in Supplementary Table S5.

\subsection{Host Ecology and Viral Lineage Patterns}

\begin{figure}[ht]
	\centering
	\includegraphics[width=0.48\textwidth]{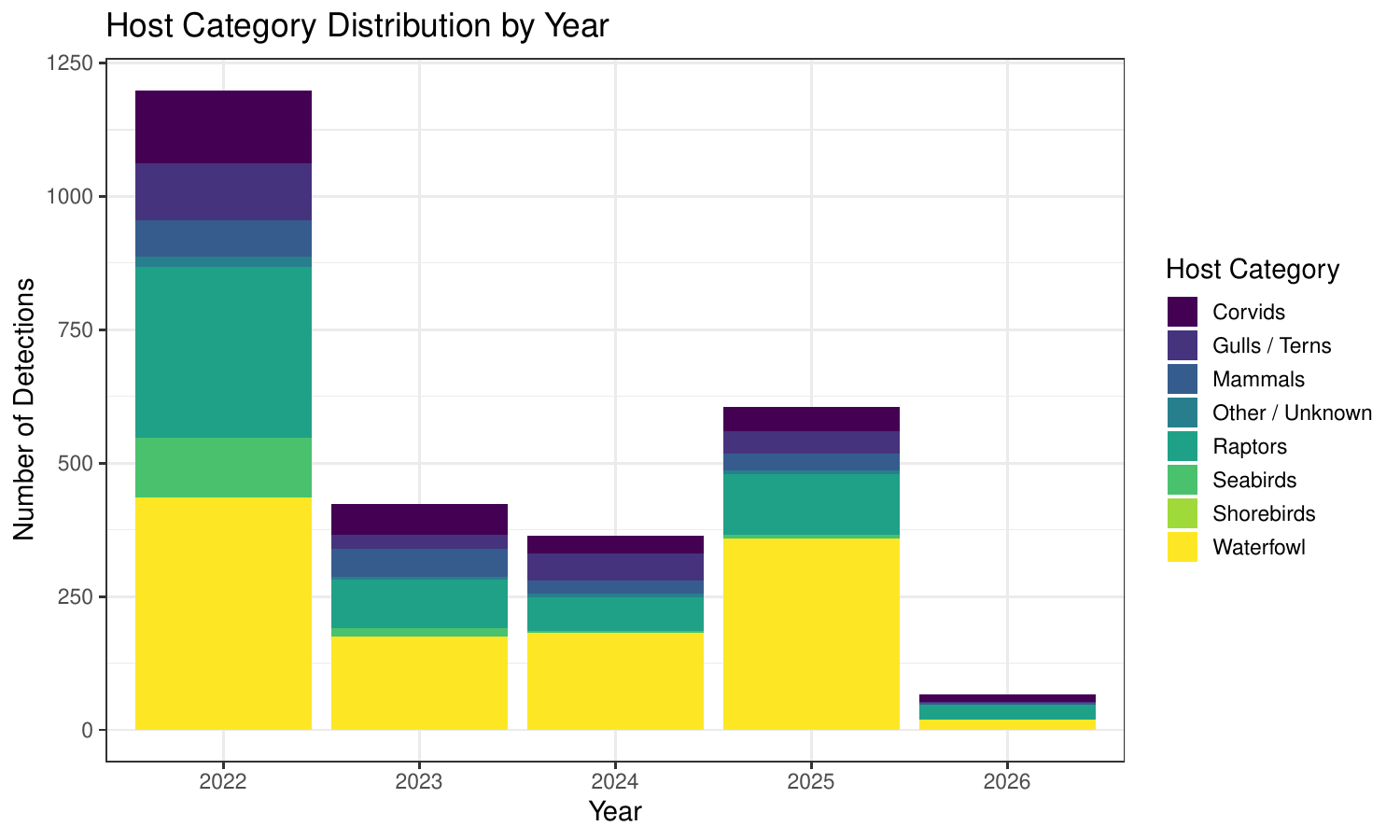}
	\includegraphics[width=0.48\textwidth]{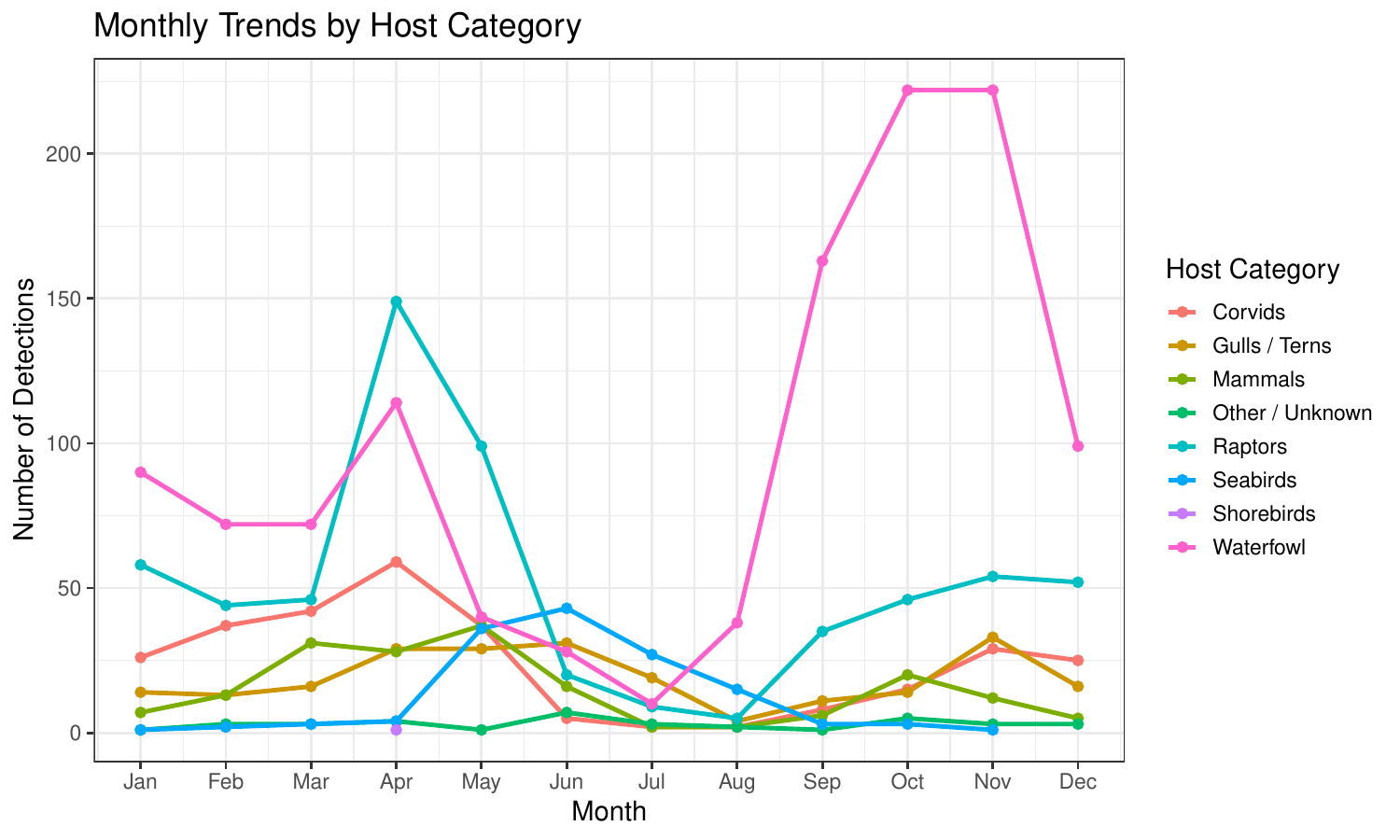}
	\caption{
		Host-category patterns of HPAI A(H5N1) detections in Canadian wildlife. Left: annual detection counts by host category from 2022 to 2026. Right: monthly detection trends by host category aggregated across the study period. Waterfowl and raptors contributed substantially to detection patterns, while corvids, gulls/terns, seabirds, and mammals contributed smaller but epidemiologically relevant components.
	}
	\label{fig:host_ecology}
\end{figure}

Host-category analysis showed that HPAI detections were concentrated among a limited number of ecological host groups (Figure~\ref{fig:host_ecology}). Waterfowl represented the largest host category, with 1,170 detections, accounting for 44.0\% of the final dataset. Raptors were the second most frequent host category, with 617 detections, representing 23.2\% of detections. Corvids accounted for 287 detections, followed by gulls/terns with 229 detections, mammals with 179 detections, and seabirds with 138 detections.


Lineage distribution was dominated by reassortment Eurasian and North American viruses. Reassortment Eu\&Na lineage accounted for 2,148 detections, representing 80.8\% of all records. Fully Eurasian lineage accounted for 494 detections, representing 18.6\%, while lineage was unknown for 15 detections, representing 0.6\% of the dataset. The full host-category and lineage distributions are provided in Supplementary Tables S3 and S4. Species-level yearly and monthly detection trends are provided in Supplementary Figures S11 and S12, and province-level host-category variation is shown in Supplementary Figure S10.

\subsection{Risk-Factor Modelling Dataset}

The zero-filled analytical modelling dataset contained 149,760 province--time--host--lineage strata. Across these strata, there were 2,657 total detections. The mean detection count was 0.018 and the variance was 0.133, giving a variance-to-mean ratio of 7.47. This indicated substantial overdispersion and supported the use of a Negative Binomial modelling framework rather than a simple Poisson model.

The modelling dataset was highly zero-heavy, with 148,693 zero-count strata, representing 99.29\% of rows, and 1,067 positive-count strata, representing 0.71\% of rows. The final modelling dataset included 13 provinces/territories, 58 species groups, eight bird categories, and three lineage groups. A summary of the zero-filled modelling dataset is provided in Supplementary Table S6.


\subsection{Risk Factors Associated with HPAI Detection Counts}

Model comparison supported the Negative Binomial season model as the primary risk-factor model. This model had the lowest AIC (13,195.84), followed closely by the zero-inflated Negative Binomial season model (AIC = 13,197.84). The Negative Binomial migration model had a higher AIC (13,257.25), and the Poisson model had the poorest fit (AIC = 20,609.60). Full model comparison results are provided in Supplementary Table S7.


Simulation-based residual diagnostics supported the adequacy of the Negative Binomial season model. The model showed no evidence of problematic overdispersion (dispersion test p = 0.330) and no evidence of residual zero inflation (zero-inflation test p = 0.860). The zero-inflated model did not improve fit relative to the simpler Negative Binomial model. Simulation-based diagnostic test results are summarised in Supplementary Table S8, and residual diagnostic plots are provided in Supplementary Figures S16 and S17.

\begin{table}[ht!]
	\centering
	\caption{Key fixed-effect estimates from the main Negative Binomial generalised linear mixed model for HPAI A(H5N1) detection counts in Canadian wildlife.}
	\label{tab:irr_main_model}
	\begin{tabular}{lrrr}
		\toprule
		\textbf{Predictor} & \textbf{IRR} & \textbf{95\% CI} & \textbf{p-value} \\
		\midrule
		Year 2023 vs 2022 & 0.24 & 0.19--0.30 & $<0.001$ \\
		Year 2024 vs 2022 & 0.20 & 0.16--0.25 & $<0.001$ \\
		Year 2025 vs 2022 & 0.28 & 0.23--0.35 & $<0.001$ \\
		Year 2026 vs 2022 & 0.03 & 0.02--0.04 & $<0.001$ \\
		Autumn vs Winter & 1.51 & 1.22--1.88 & $<0.001$ \\
		Spring vs Winter & 1.39 & 1.12--1.73 & 0.003 \\
		Summer vs Winter & 0.61 & 0.48--0.79 & $<0.001$ \\
		Shorebirds vs Waterfowl & 0.04 & 0.01--0.35 & 0.003 \\
		Fully Eurasian lineage & 40.39 & 23.30--70.00 & $<0.001$ \\
		Reassortment Eu\&Na lineage & 129.80 & 75.37--223.52 & $<0.001$ \\
		\bottomrule
	\end{tabular}
\end{table}

In the final Negative Binomial season model, detection counts were significantly lower in all years after 2022. Compared with 2022, the IRR was 0.24 for 2023, 0.20 for 2024, 0.28 for 2025, and 0.03 for 2026. Compared with winter, autumn had a significantly higher expected detection count (IRR = 1.51, 95\% CI: 1.22--1.88), and spring also had a higher expected detection count (IRR = 1.39, 95\% CI: 1.12--1.73). Summer had a significantly lower expected detection count than winter (IRR = 0.61, 95\% CI: 0.48--0.79).

\begin{figure}[ht]
	\centering
	\begin{subfigure}[t]{0.32\textwidth}
		\centering
		\includegraphics[width=\textwidth]{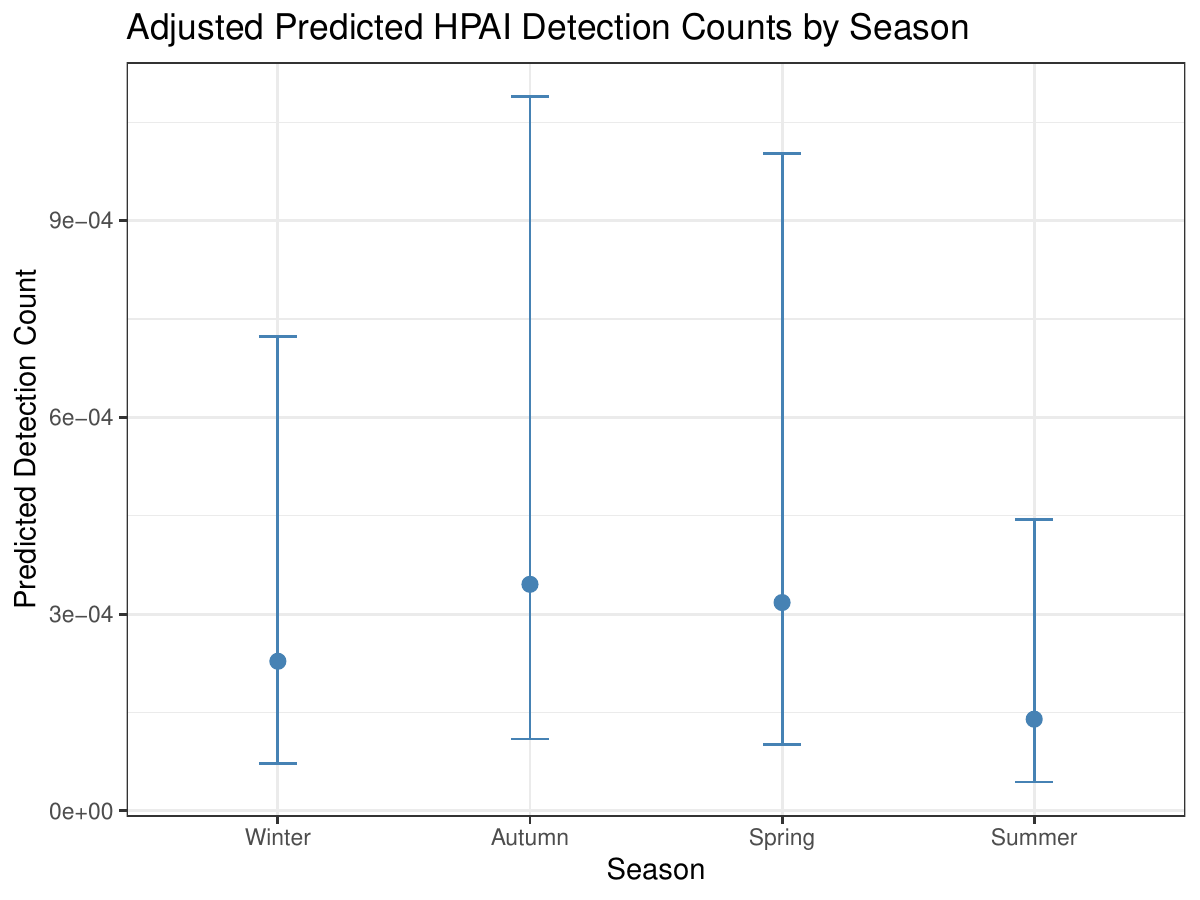}
		\caption{Season}
		\label{fig:model_predictions_season}
	\end{subfigure}
	\hfill
	\begin{subfigure}[t]{0.32\textwidth}
		\centering
		\includegraphics[width=\textwidth]{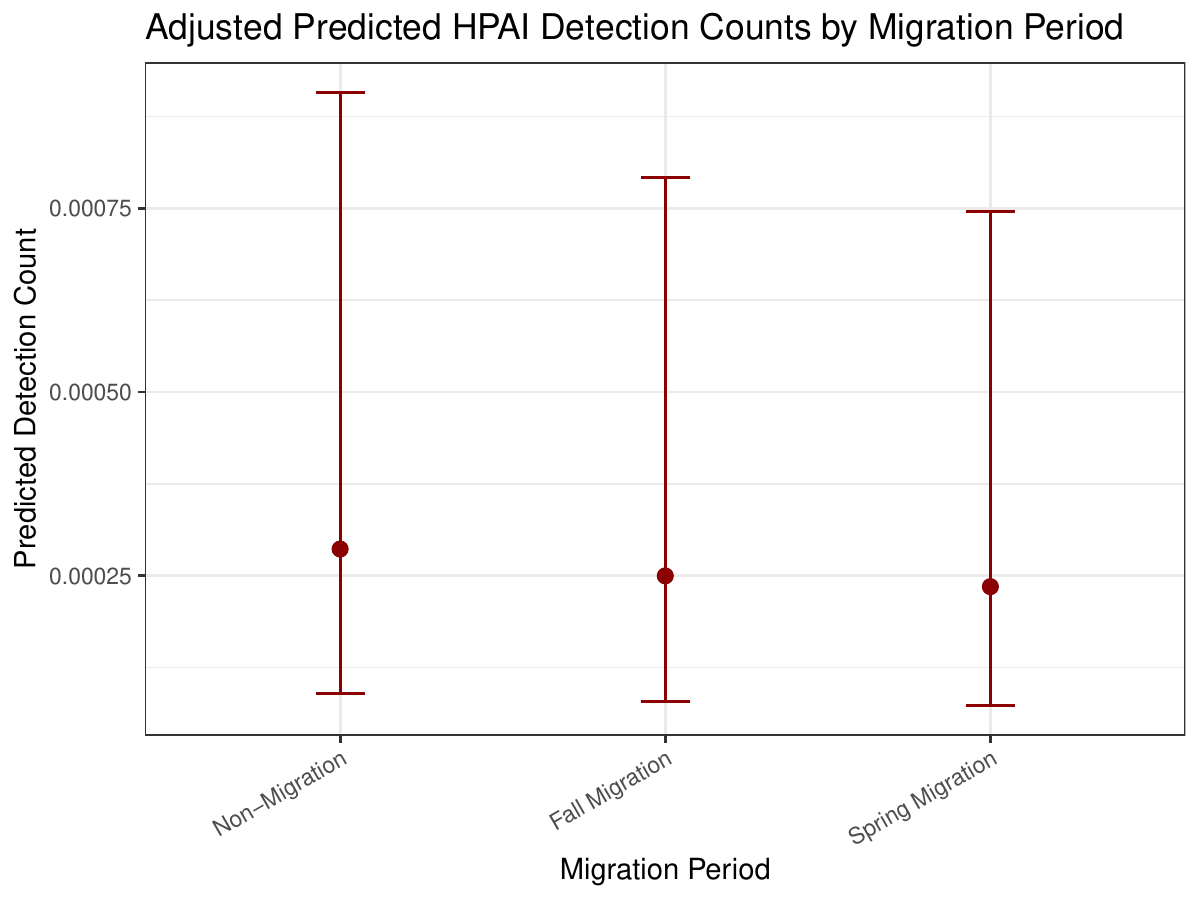}
		\caption{Migration period}
		\label{fig:model_predictions_migration}
	\end{subfigure}
	\hfill
	\begin{subfigure}[t]{0.32\textwidth}
		\centering
		\includegraphics[width=\textwidth]{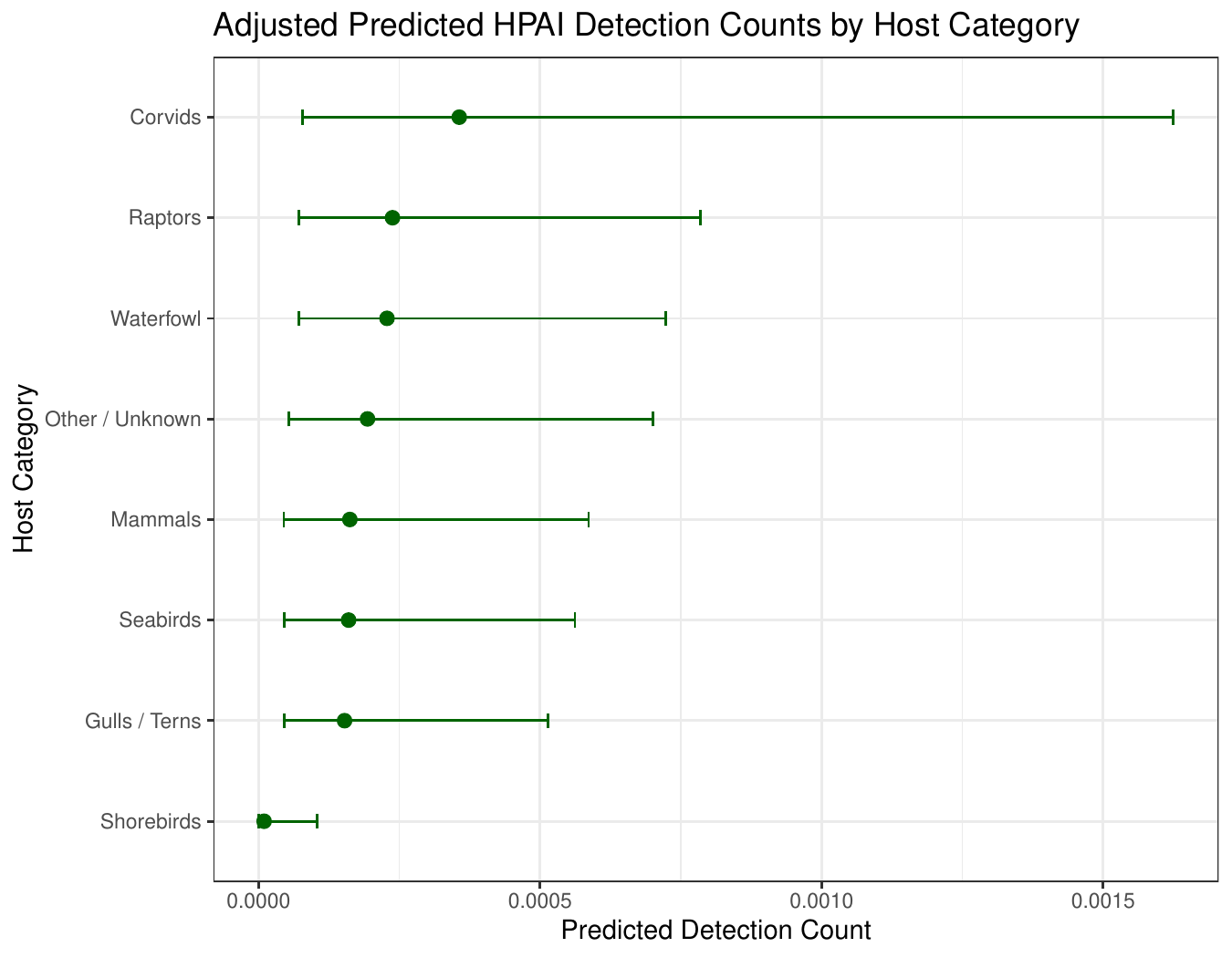}
		\caption{Host category}
		\label{fig:model_predictions_host}
	\end{subfigure}
	\caption{
		Adjusted predicted HPAI A(H5N1) detection counts from the Negative Binomial generalised linear mixed model. Panels show adjusted predictions by season, migration period, and host category. Predictions account for fixed effects and province- and species-level random intercepts.
	}
	\label{fig:model_predictions}
\end{figure}

Lineage was strongly associated with detection counts. Fully Eurasian lineage had a substantially higher expected detection count (IRR = 40.39, 95\% CI: 23.30--70.00), while Reassortment Eu\&Na lineage had the highest expected detection count (IRR = 129.80, 95\% CI: 75.37--223.52). Adjusted model predictions are shown in Figure~\ref{fig:model_predictions}.


\section{Discussion}

This study provides a national One Health analysis of highly pathogenic avian influenza A(H5N1) detections in Canadian wildlife from 2022 to early 2026. The findings show that detection patterns were consistently structured by temporal dynamics, geographic variation, host ecology, and viral lineage, highlighting the complex ecological and epidemiological drivers of HPAI transmission.

The temporal distribution of detections was strongly influenced by the initial emergence and subsequent progression of clade 2.3.4.4b HPAI A(H5N1) in North America. The highest detection burden occurred in 2022, corresponding to the early phase following the late-2021 introduction into eastern Canada. Although detection counts declined in 2023 and 2024, the resurgence observed in 2025 suggests continued viral circulation in wildlife populations, supporting the need for sustained long-term surveillance.

Seasonality was a key determinant of detection intensity. Detection counts were significantly higher in autumn and spring compared with winter, while summer exhibited reduced detection activity. These patterns are consistent with known avian influenza dynamics influenced by migration, seasonal aggregation, environmental persistence, and changing host-contact structures. However, the substantial number of detections observed during non-migration periods indicates that transmission is not restricted to migratory events alone. Consequently, surveillance efforts should be maintained year-round, with targeted intensification during high-risk seasonal windows.

Spatial analyses revealed marked geographic heterogeneity in detection burden. Ontario, Alberta, and British Columbia were consistently identified as high-burden regions, while additional provinces showed moderate to high detection levels. These patterns likely reflect a combination of ecological factors, host density, surveillance intensity, and reporting capacity. Importantly, detection hotspots should be interpreted as areas of high observed detection burden rather than definitive estimates of underlying infection prevalence, as surveillance-based data are inherently influenced by reporting processes and sampling effort.

Host ecology played a central role in structuring detection patterns. Waterfowl represented the dominant host group, consistent with their established role as natural reservoirs and dispersal agents of avian influenza viruses. Raptors, corvids, gulls/terns, seabirds, and mammals also contributed meaningfully to detections, reflecting multiple ecological transmission pathways, including environmental exposure, predation, and scavenging. The involvement of seabirds and mammalian wildlife has important conservation and One Health implications, particularly given the documented large-scale mortality events in seabirds during the early phase of the outbreak. This conservation relevance is especially important for vulnerable wildlife populations, where repeated HPAI-associated mortality events could affect population recovery, breeding success, and long-term ecosystem resilience.

Viral lineage was one of the strongest predictors of detection intensity. Reassortant Eurasian–North American viral lineages dominated the dataset and were associated with substantially higher detection counts compared with other lineage groups. These findings are consistent with genomic evidence indicating multiple introductions and extensive reassortment events following the emergence of clade 2.3.4.4b in North America. However, the large magnitude of lineage-associated incidence rate ratios should be interpreted relative to the reference lineage category and may partly reflect sparse counts in less frequent lineage groups.

The modelling results support the use of a Negative Binomial generalised linear mixed modelling framework for analysing wildlife surveillance data. The dataset exhibited substantial overdispersion and a high proportion of zero-count strata, making standard Poisson approaches unsuitable. Year, season, and lineage emerged as the main drivers of detection intensity, reinforcing the importance of integrating temporal, ecological, and virological factors in risk assessment models.

This study has several limitations. The analysis was based on surveillance detections rather than true infection prevalence and is therefore subject to biases associated with reporting effort, carcass detection probability, wildlife density, laboratory submission practices, and regional surveillance capacity. Mortality-associated detections may overrepresent severe infections, while mild or subclinical infections are likely underrepresented. Additionally, province-level aggregation may mask finer-scale ecological variability, and key environmental covariates such as habitat, climate, and poultry interface dynamics were not included. Host-category classification also introduced partial overlap with host-type variables, limiting estimation of certain model parameters.

Despite these limitations, the findings provide important insights for surveillance and risk management. A risk-based One Health surveillance approach should prioritise high-burden regions, migration-associated periods, waterbird and scavenger-associated host groups, reassortant viral lineages, and continued monitoring of mammalian wildlife. These results provide a practical framework for designing targeted surveillance strategies in regions experiencing ongoing HPAI expansion. Future research should aim to integrate wildlife surveillance with environmental, genomic, poultry, and public health data to strengthen early warning systems and improve preparedness for HPAI A(H5N1) and other emerging zoonotic threats.

\section{Conclusion}

Canadian wildlife HPAI A(H5N1) detections from 2022 to early 2026 showed clear temporal, spatial, host-associated, and virological structure. Detections were concentrated in high-burden provinces, were dominated by avian hosts, particularly waterfowl and raptors, and were higher during autumn and spring. Reassortant Eurasian–North American viral lineages were the dominant lineage group and were strongly associated with detection intensity. These findings support risk-based One Health surveillance that prioritises high-burden regions, migration-associated periods, waterbird and scavenger-associated host groups, reassortant viral lineages, and mammalian wildlife monitoring. Future surveillance should integrate wildlife, domestic animal, environmental, genomic, and public health data to improve early warning and preparedness for HPAI A(H5N1) and other emerging zoonotic threats.

\vspace{2cm}

	\begin{center}
	{\Large \textbf{Supplementary Material}}
\end{center}

\vspace{1em}

\setcounter{table}{0}
\setcounter{figure}{0}


\section*{Supplementary Tables}


\begin{table}[H]
	\centering
	\caption{Annual distribution of HPAI A(H5N1) wildlife detections in Canada.}
	\label{tab:supp_annual_detections}
	\begin{tabular}{lrr}
		\toprule
		\textbf{Year} & \textbf{Detection Count} & \textbf{Percentage} \\
		\midrule
		2022 & 1,199 & 45.1 \\
		2023 & 423 & 15.9 \\
		2024 & 363 & 13.7 \\
		2025 & 605 & 22.8 \\
		2026 & 67 & 2.5 \\
		\midrule
		\textbf{Total} & \textbf{2,657} & \textbf{100.0} \\
		\bottomrule
	\end{tabular}
\end{table}
\FloatBarrier

\vspace{1.5em}

\begin{table}[H]
	\centering
	\caption{Seasonal and migration-period distribution of HPAI A(H5N1) wildlife detections in Canada.}
	\label{tab:supp_season_migration_distribution}
	\begin{tabular}{llrr}
		\toprule
		\textbf{Classification} & \textbf{Category} & \textbf{Detection Count} & \textbf{Percentage} \\
		\midrule
		Season & Autumn & 906 & 34.1 \\
		Season & Spring & 880 & 33.1 \\
		Season & Winter & 581 & 21.9 \\
		Season & Summer & 290 & 10.9 \\
		\midrule
		Migration period & Fall migration & 906 & 34.1 \\
		Migration period & Spring migration & 880 & 33.1 \\
		Migration period & Non-migration & 871 & 32.8 \\
		\bottomrule
	\end{tabular}
\end{table}
\FloatBarrier

\vspace{1.5em}

\begin{table}[H]
	\centering
	\caption{Host-category distribution of HPAI A(H5N1) wildlife detections in Canada.}
	\label{tab:supp_host_category_distribution}
	\begin{tabular}{lrr}
		\toprule
		\textbf{Host Category} & \textbf{Detection Count} & \textbf{Percentage} \\
		\midrule
		Waterfowl & 1,170 & 44.0 \\
		Raptors & 617 & 23.2 \\
		Corvids & 287 & 10.8 \\
		Gulls/Terns & 229 & 8.6 \\
		Mammals & 179 & 6.7 \\
		Seabirds & 138 & 5.2 \\
		Other/Unknown & 36 & 1.4 \\
		Shorebirds & 1 & 0.0 \\
		\midrule
		\textbf{Total} & \textbf{2,657} & \textbf{100.0} \\
		\bottomrule
	\end{tabular}
\end{table}
\FloatBarrier

\vspace{1.5em}

\begin{table}[H]
	\centering
	\caption{Lineage distribution of HPAI A(H5N1) wildlife detections in Canada.}
	\label{tab:supp_lineage_distribution}
	\begin{tabular}{lrr}
		\toprule
		\textbf{Lineage} & \textbf{Detection Count} & \textbf{Percentage} \\
		\midrule
		Reassortment Eu\&Na & 2,148 & 80.8 \\
		Fully Eurasian & 494 & 18.6 \\
		Unknown & 15 & 0.6 \\
		\midrule
		\textbf{Total} & \textbf{2,657} & \textbf{100.0} \\
		\bottomrule
	\end{tabular}
\end{table}
\FloatBarrier

\vspace{1.5em}

\begin{table}[H]
	\centering
	\caption{Significant Local Moran's I results for province-level HPAI A(H5N1) wildlife detection counts.}
	\label{tab:supp_lisa_significant}
	\begin{tabular}{lrrrrl}
		\toprule
		\textbf{Province} & \textbf{Detection Count} & \textbf{Local I} & \textbf{z-score} & \textbf{p-value} & \textbf{LISA Cluster} \\
		\midrule
		Saskatchewan & 249 & 0.2697 & 2.5109 & 0.0120 & High-High \\
		Manitoba & 167 & -0.2495 & -2.6800 & 0.0074 & Low-High \\
		\bottomrule
	\end{tabular}
\end{table}
\FloatBarrier

\vspace{1.5em}

\begin{table}[H]
	\centering
	\caption{Summary of the zero-filled analytical dataset used for risk-factor modelling..}
	\label{tab:supp_model_data_summary}
	\begin{tabular}{lr}
		\toprule
		\textbf{Characteristic} & \textbf{Value} \\
		\midrule
		Number of modelling strata & 149,760 \\
		Total detections & 2,657 \\
		Mean detection count & 0.018 \\
		Variance of detection count & 0.133 \\
		Variance-to-mean ratio & 7.47 \\
		Zero-count strata & 148,693 \\
		Zero-count strata, \% & 99.29 \\
		Positive-count strata & 1,067 \\
		Positive-count strata, \% & 0.71 \\
		Number of provinces/territories & 13 \\
		Number of species groups & 58 \\
		Number of bird categories & 8 \\
		Number of lineage groups & 3 \\
		\bottomrule
	\end{tabular}
\end{table}
\FloatBarrier

\vspace{1.5em}

\begin{table}[H]
	\centering
	\caption{Model comparison for HPAI A(H5N1) detection-count models.}
	\label{tab:supp_model_comparison}
	\begin{tabular}{lrr}
		\toprule
		\textbf{Model} & \textbf{df} & \textbf{AIC} \\
		\midrule
		Negative Binomial GLMM: season model & 20 & 13,195.84 \\
		Zero-inflated Negative Binomial GLMM: season model & 21 & 13,197.84 \\
		Negative Binomial GLMM: migration model & 19 & 13,257.25 \\
		Poisson GLMM: season model & 19 & 20,609.60 \\
		\bottomrule
	\end{tabular}
\end{table}
\FloatBarrier

\vspace{1.5em}

\begin{table}[H]
	\centering
	\caption{Simulation-based residual diagnostic tests for candidate detection-count models.}
	\label{tab:supp_model_diagnostics}
	\begin{tabular}{lrr}
		\toprule
		\textbf{Model} & \textbf{Dispersion p-value} & \textbf{Zero-inflation p-value} \\
		\midrule
		Poisson season model & 0.348 & 0.074 \\
		Negative Binomial season model & 0.330 & 0.860 \\
		Negative Binomial migration model & 0.384 & 0.914 \\
		Zero-inflated Negative Binomial season model & 0.326 & 0.838 \\
		\bottomrule
	\end{tabular}
\end{table}
\FloatBarrier


\clearpage
\section*{Supplementary Figures}


\begin{figure}[H]
	\centering
	\includegraphics[width=0.85\textwidth]{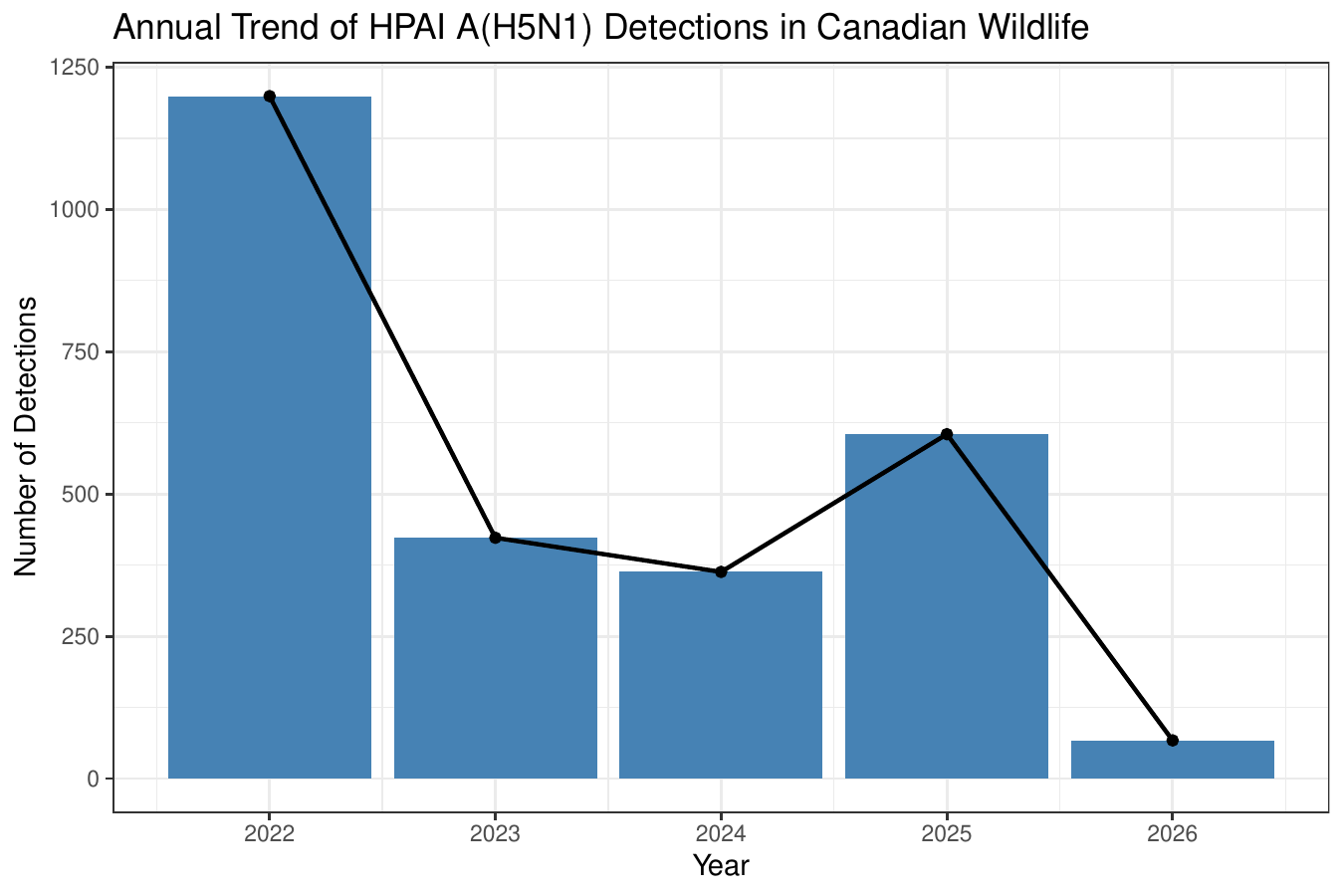}
	\caption{Annual trend of HPAI A(H5N1) detections in Canadian wildlife, 2022--2026. The highest number of detections occurred in 2022, followed by lower counts in 2023 and 2024, increased detections in 2025, and partial-year detections in 2026.}
	\label{fig:supp_annual_trend}
\end{figure}
\FloatBarrier

\clearpage

\begin{figure}[H]
	\centering
	\includegraphics[width=0.90\textwidth]{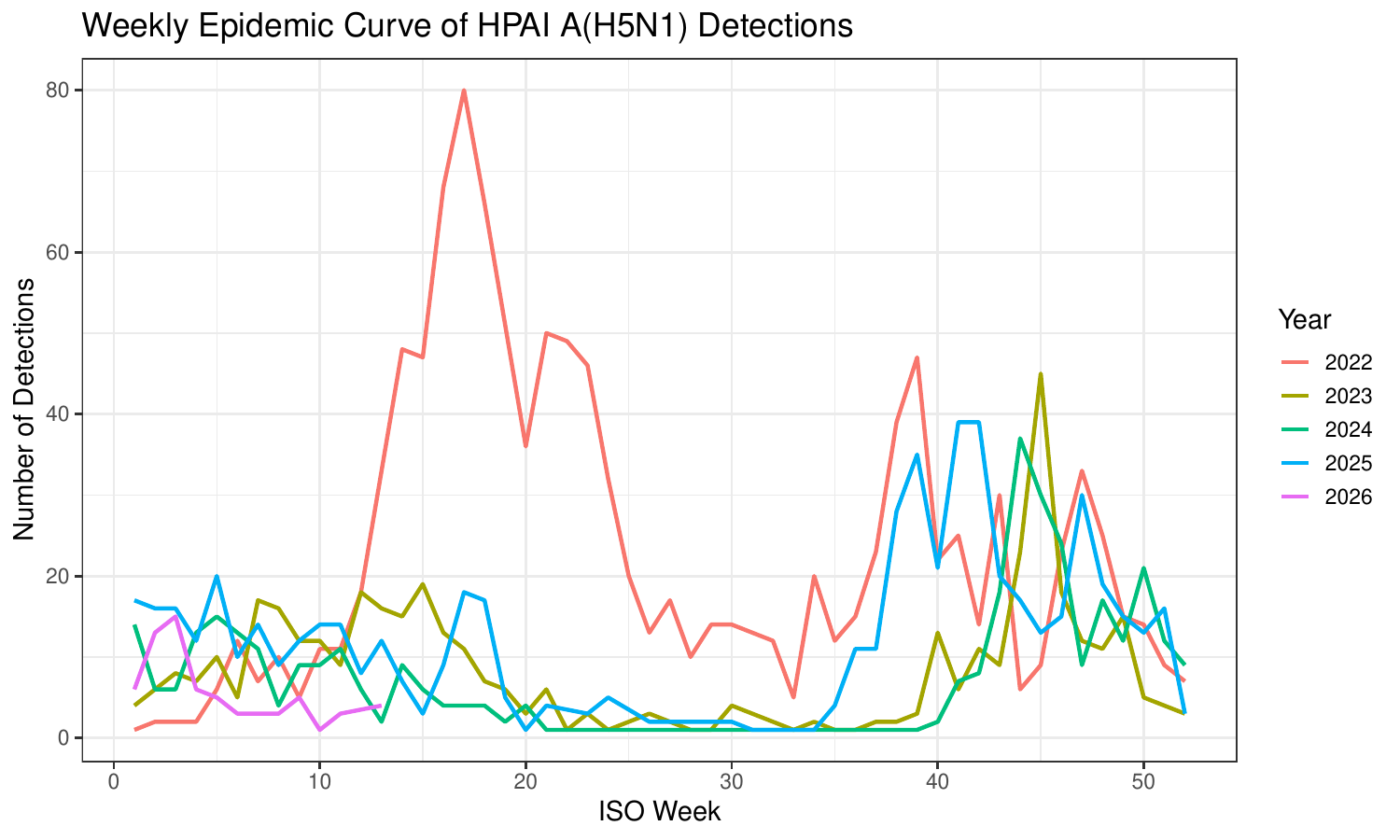}
	\caption{Weekly epidemic curve of HPAI A(H5N1) detections in Canadian wildlife by ISO week and year. Weekly counts show distinct epidemic waves and support the temporal heterogeneity observed in the monthly epidemic curve.}
	\label{fig:supp_weekly_epidemic_curve}
\end{figure}
\FloatBarrier

\clearpage

\begin{figure}[H]
	\centering
	\includegraphics[width=0.95\textwidth]{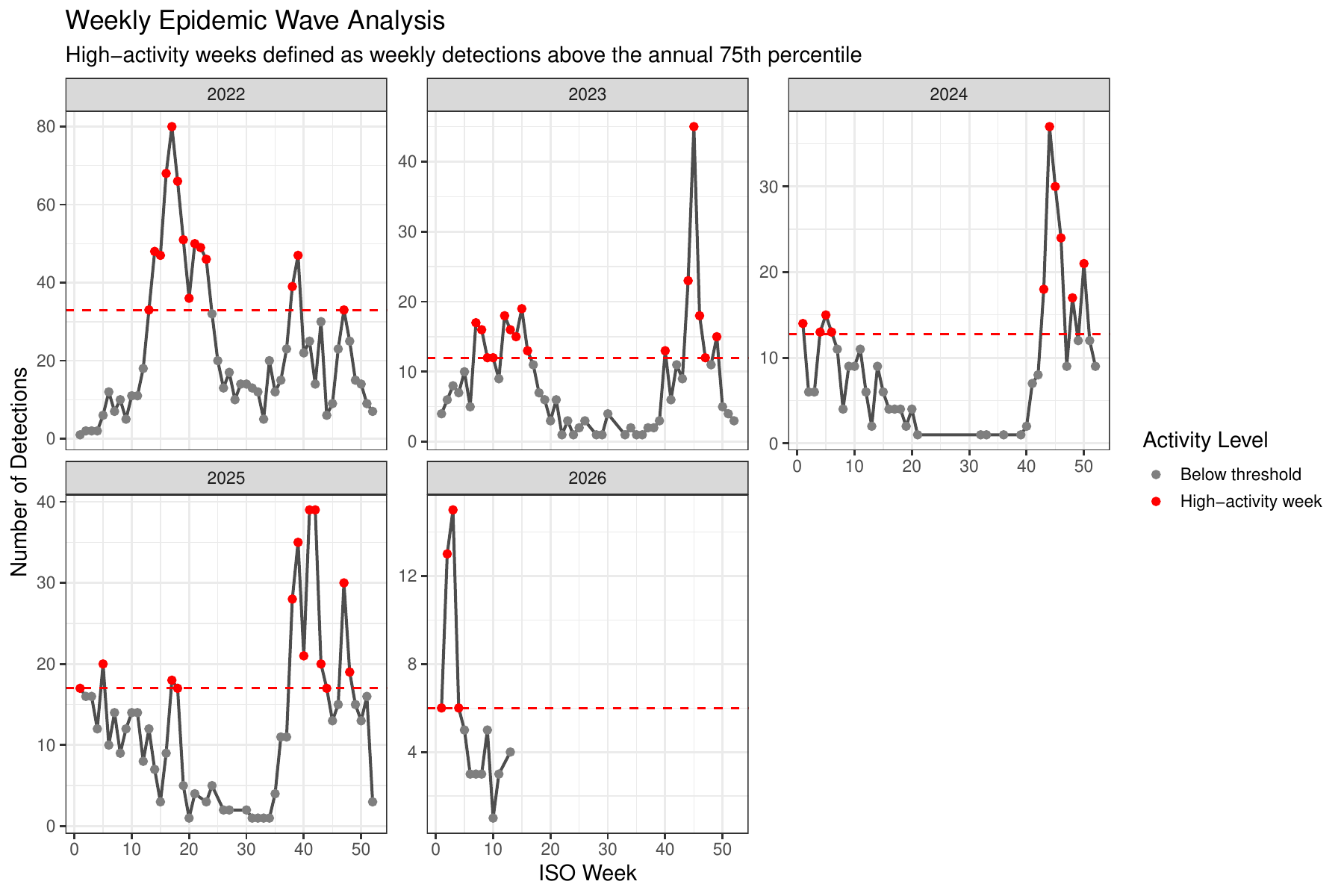}
	\caption{Weekly epidemic-wave analysis of HPAI A(H5N1) detections in Canadian wildlife. High-activity weeks were defined as weekly detection counts above the annual 75th percentile threshold. The figure shows year-specific high-activity periods across ISO weeks.}
	\label{fig:supp_weekly_wave_thresholds}
\end{figure}
\FloatBarrier


\clearpage
\begin{figure}[H]
	\centering
	\includegraphics[width=0.90\textwidth]{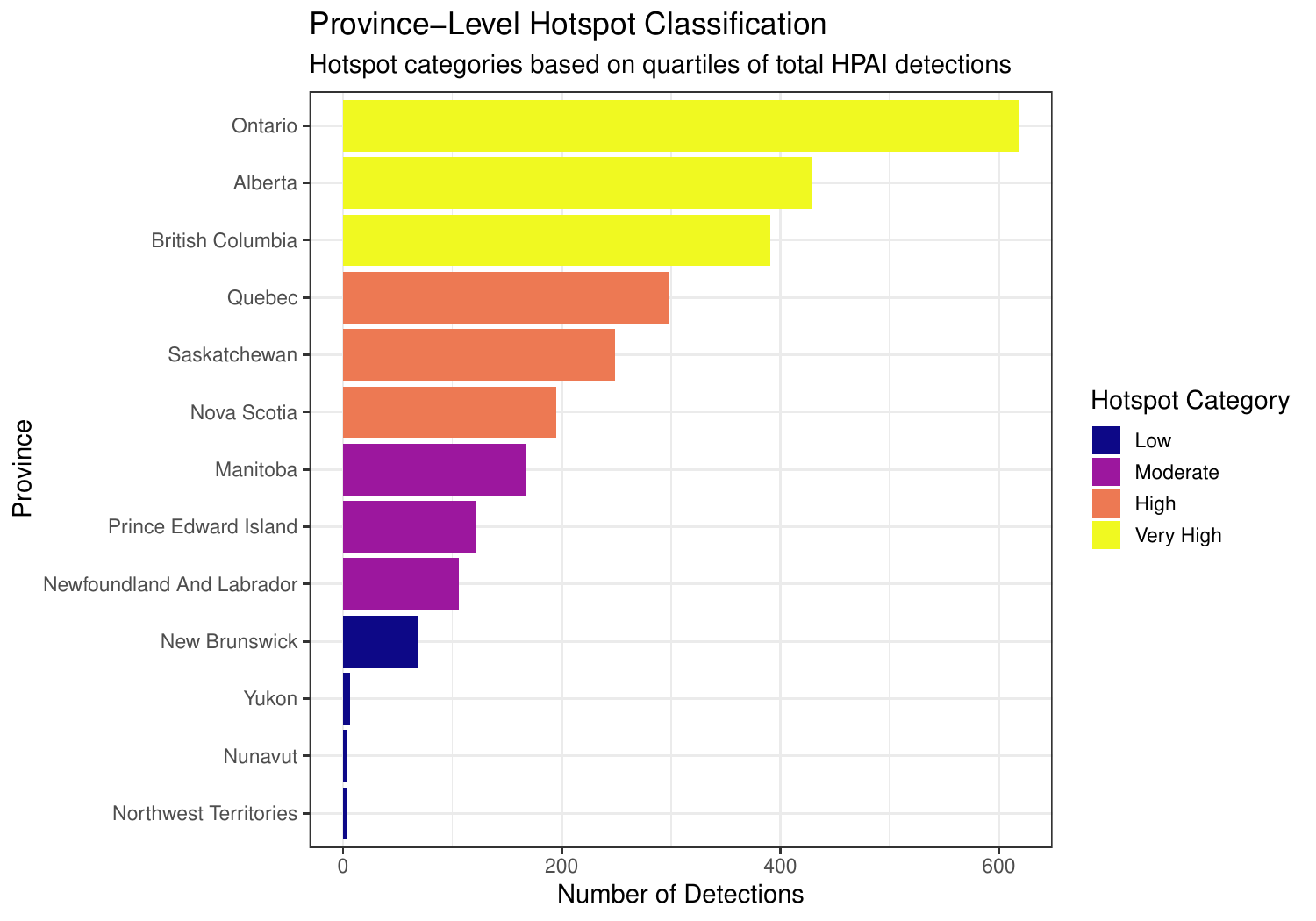}
	\caption{Province-level hotspot classification of HPAI A(H5N1) detections in Canadian wildlife. Detection counts were classified into low, moderate, high, and very high hotspot categories using quartiles of total provincial detection counts.}
	\label{fig:supp_province_hotspot_bar}
\end{figure}
\FloatBarrier


\clearpage
\begin{figure}[H]
	\centering
	\includegraphics[width=0.90\textwidth]{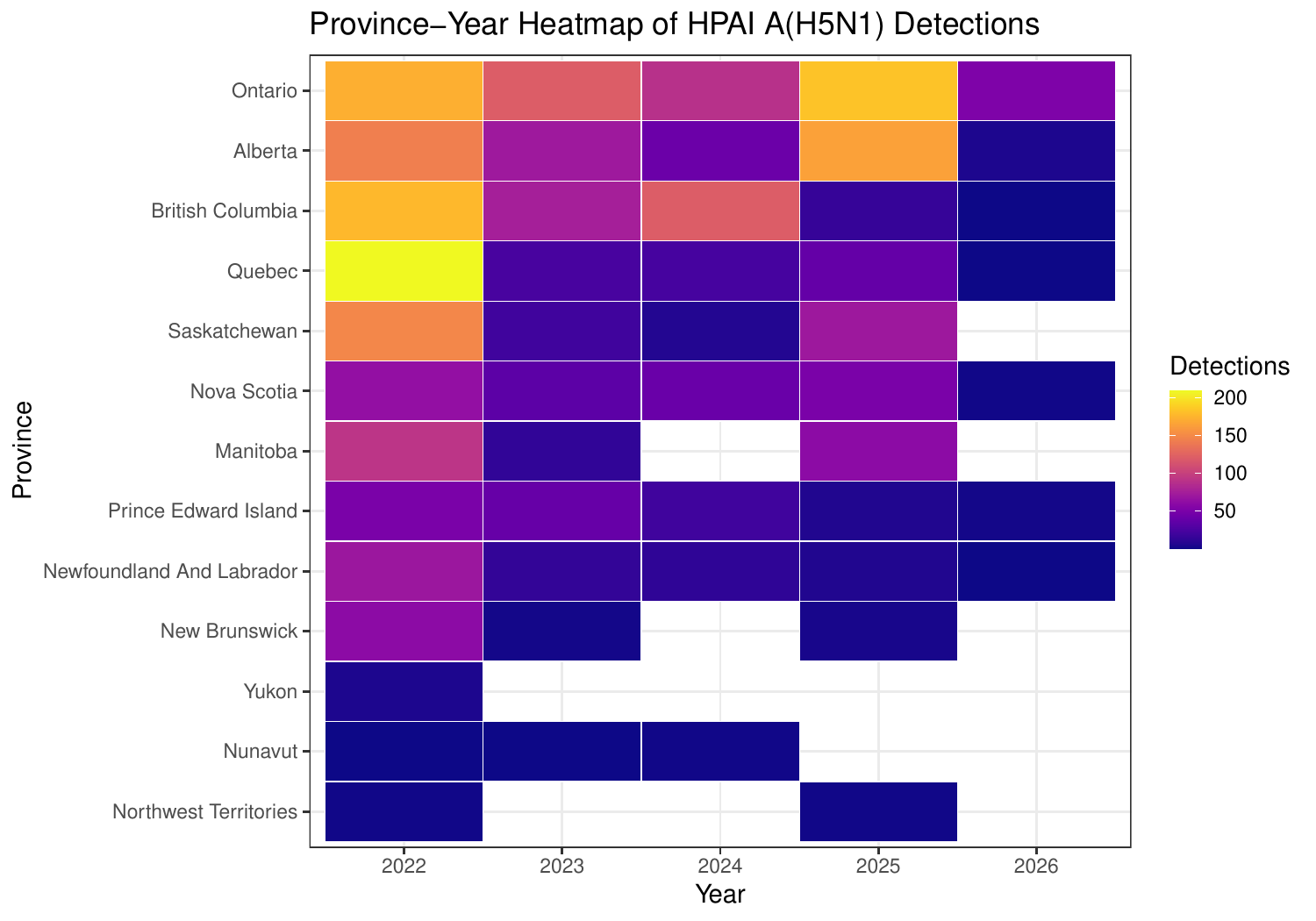}
	\caption{Province-year heatmap of HPAI A(H5N1) detections in Canadian wildlife. Detection counts were aggregated by province and year, showing temporal variation in province-level detection burden.}
	\label{fig:supp_province_year_heatmap}
\end{figure}
\FloatBarrier


\clearpage
\begin{figure}[H]
	\centering
	\includegraphics[width=0.90\textwidth]{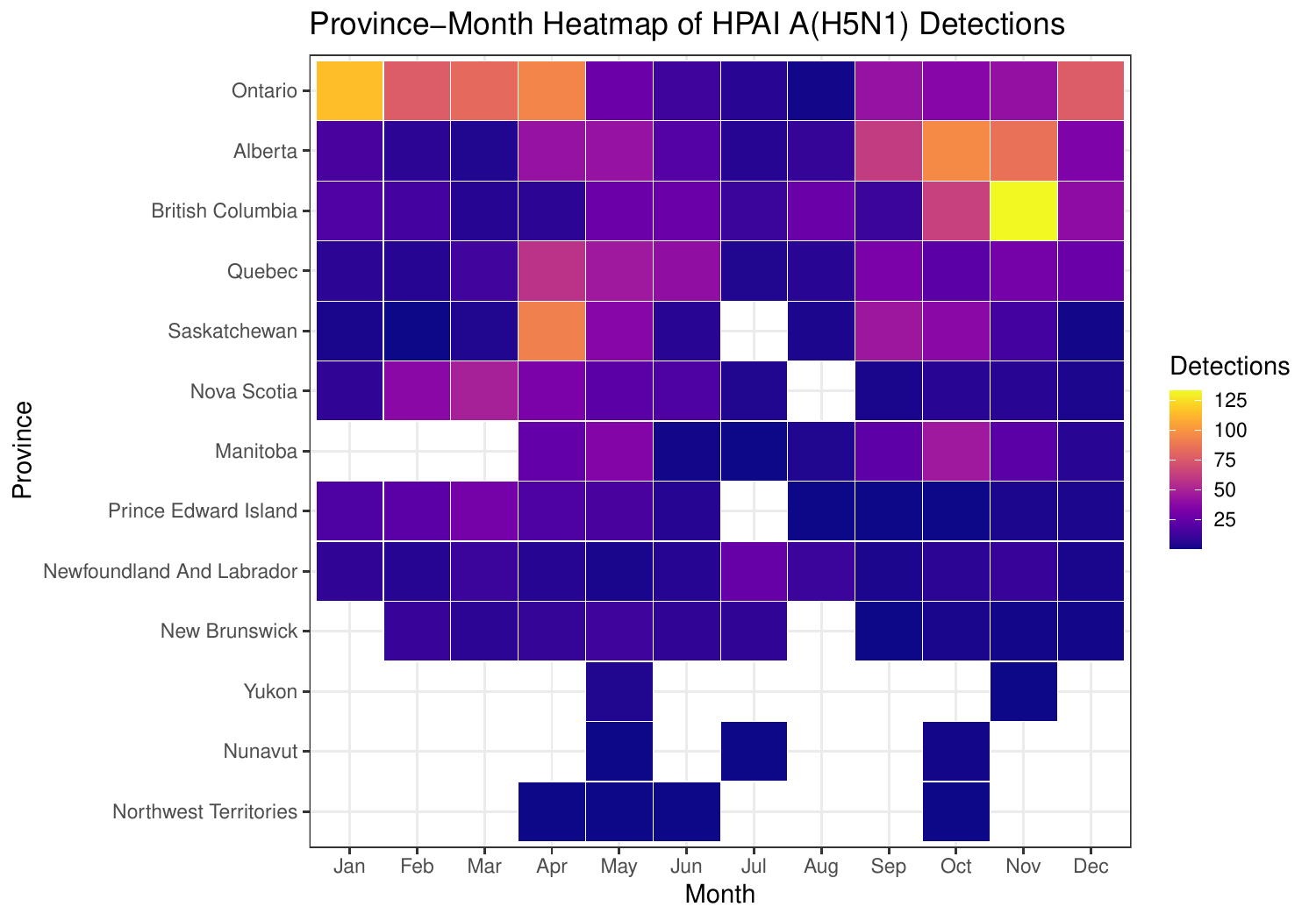}
	\caption{Province-month heatmap of HPAI A(H5N1) detections in Canadian wildlife. Detection counts were aggregated across years by province and month, showing detailed seasonal variation in province-level detection activity.}
	\label{fig:supp_province_month_heatmap}
\end{figure}
\FloatBarrier

\clearpage


\clearpage
\begin{figure}[H]
	\centering
	\includegraphics[width=0.90\textwidth]{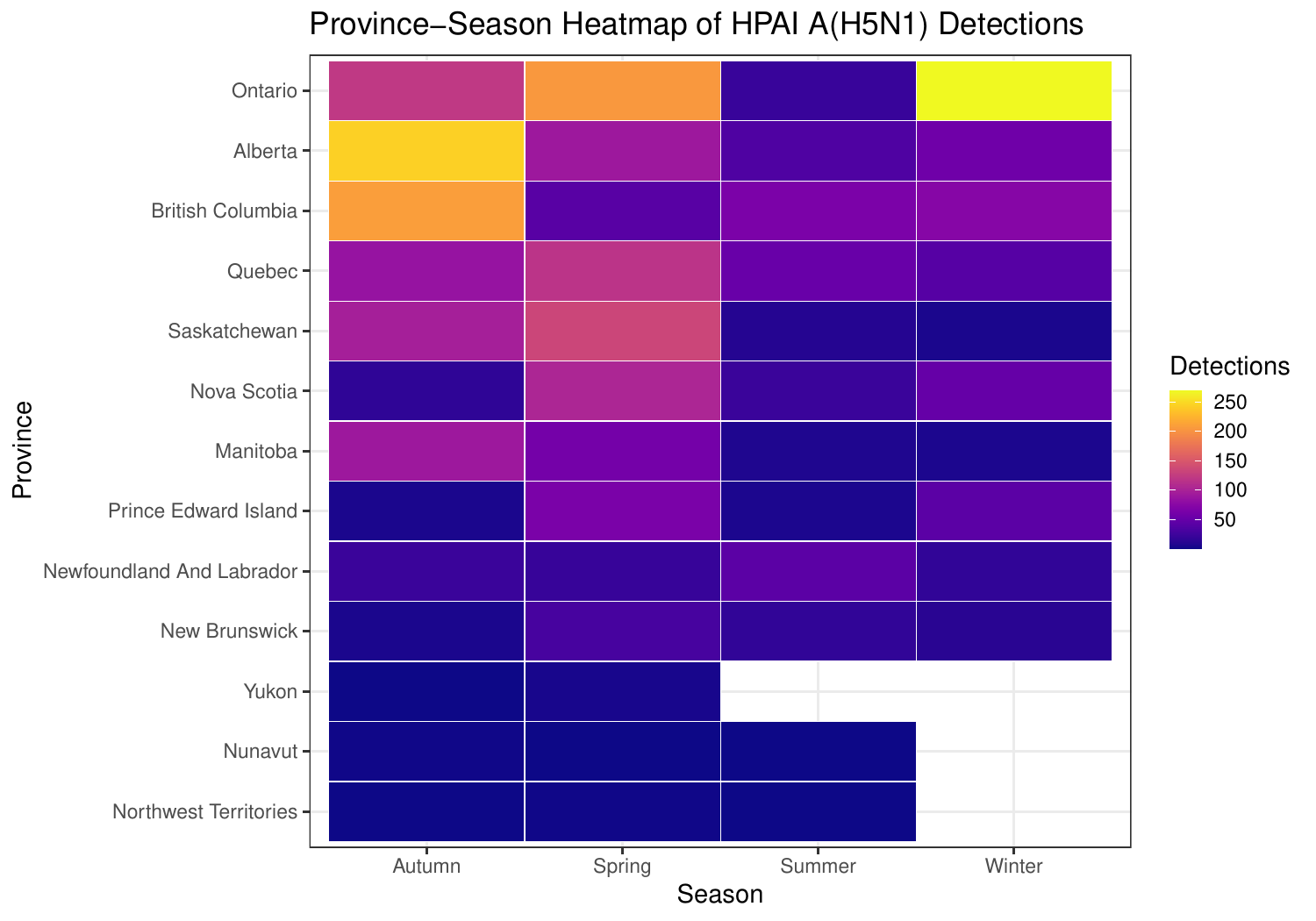}
	\caption{Province-season heatmap of HPAI A(H5N1) detections in Canadian wildlife. Detection counts were aggregated by province and season, showing seasonal heterogeneity in detection burden across provinces and territories.}
	\label{fig:supp_province_season_heatmap}
\end{figure}
\FloatBarrier

\clearpage

\begin{figure}[H]
	\centering
	\includegraphics[width=0.80\textwidth]{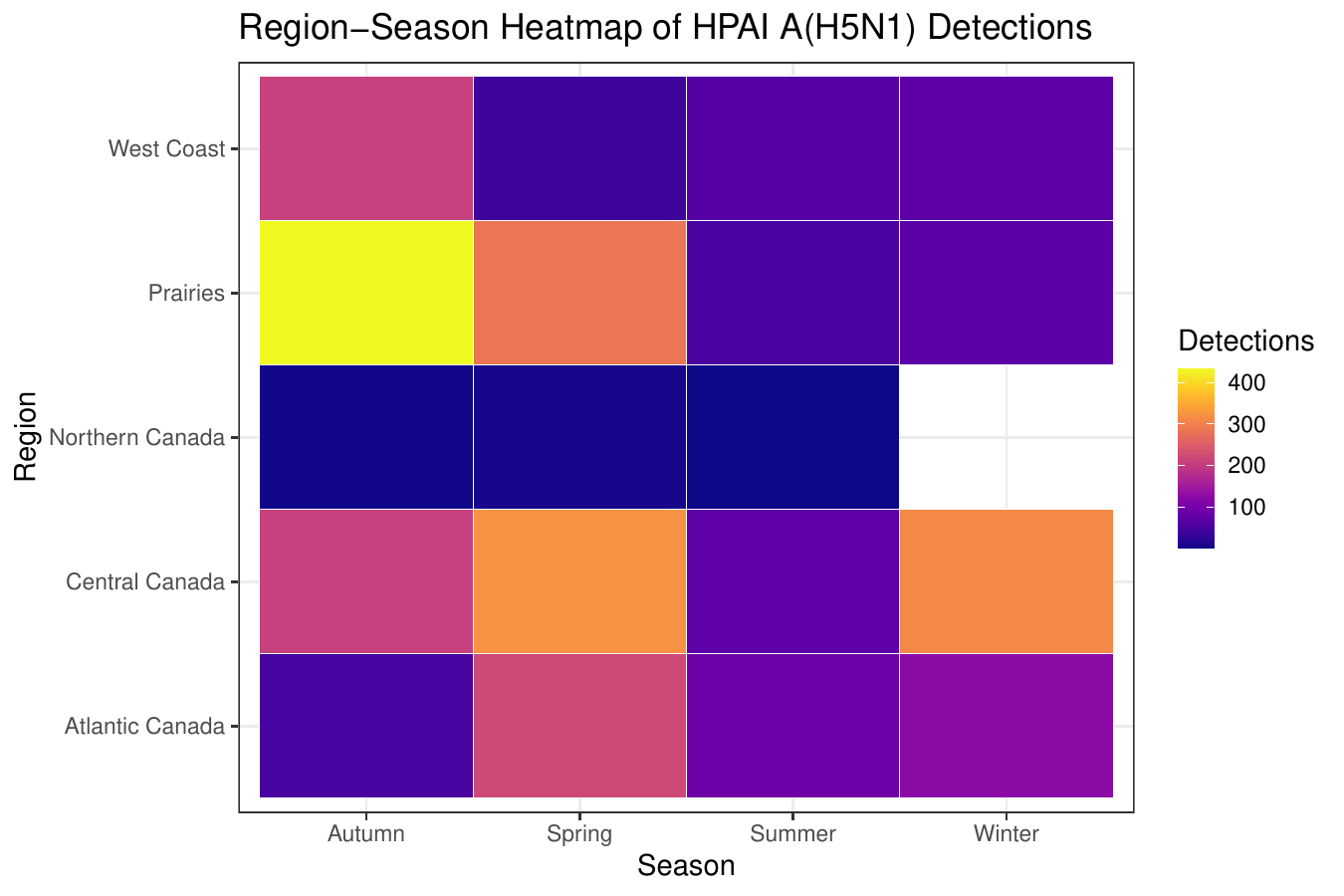}
	\caption{Region-season heatmap of HPAI A(H5N1) detections in Canadian wildlife. Detection counts were aggregated by Canadian region and season, highlighting broad spatiotemporal differences relevant to risk-based One Health surveillance.}
	\label{fig:supp_region_season_heatmap}
\end{figure}
\FloatBarrier

\clearpage

\clearpage
\begin{figure}[H]
	\centering
	\includegraphics[width=0.90\textwidth]{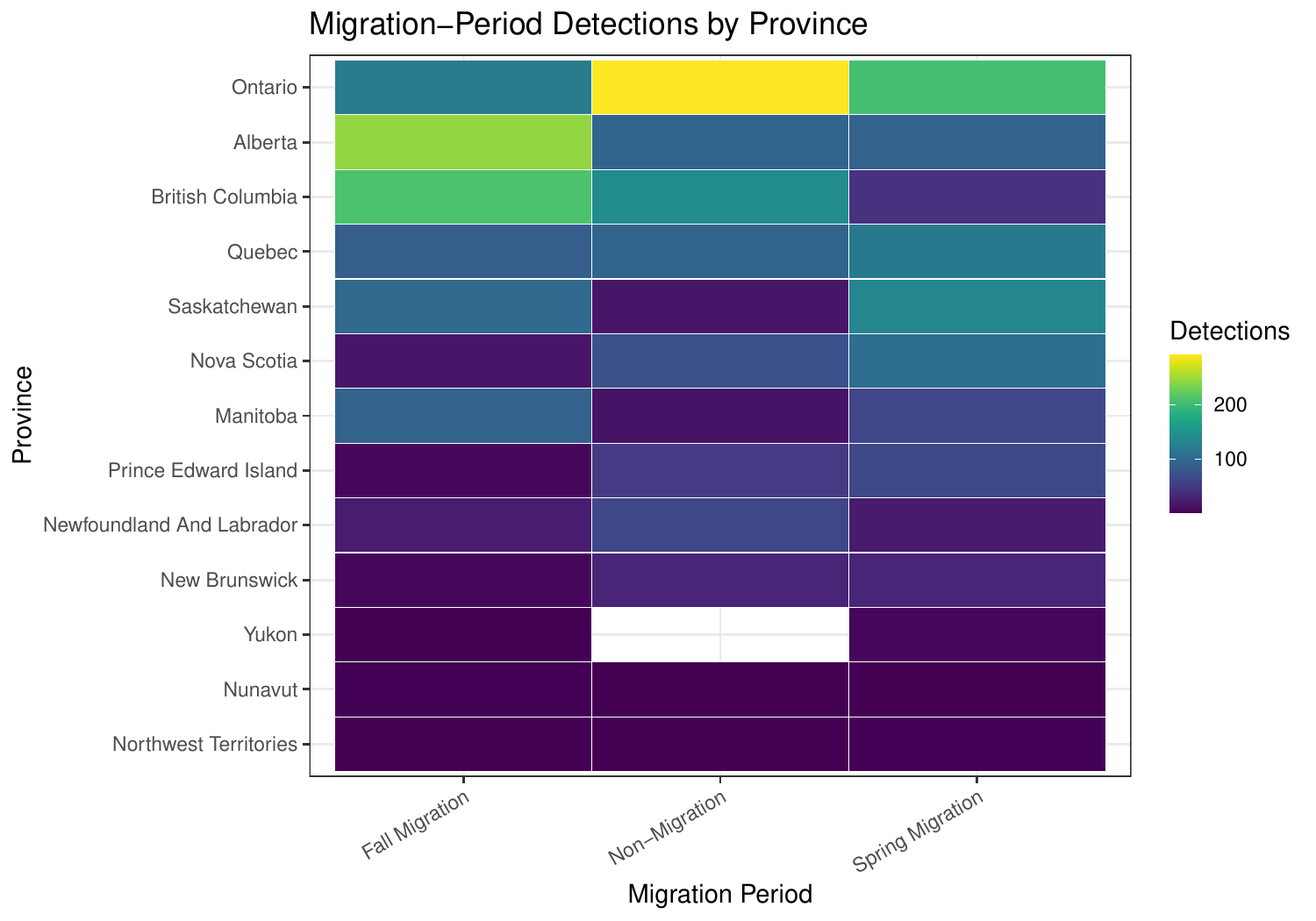}
	\caption{Province-level heatmap of HPAI A(H5N1) detections by migration period. Detection counts were aggregated by province and classified as spring migration, fall migration, or non-migration periods.}
	\label{fig:supp_province_migration_heatmap}
\end{figure}
\FloatBarrier

\clearpage

\begin{figure}[H]
	\centering
	\includegraphics[width=0.90\textwidth]{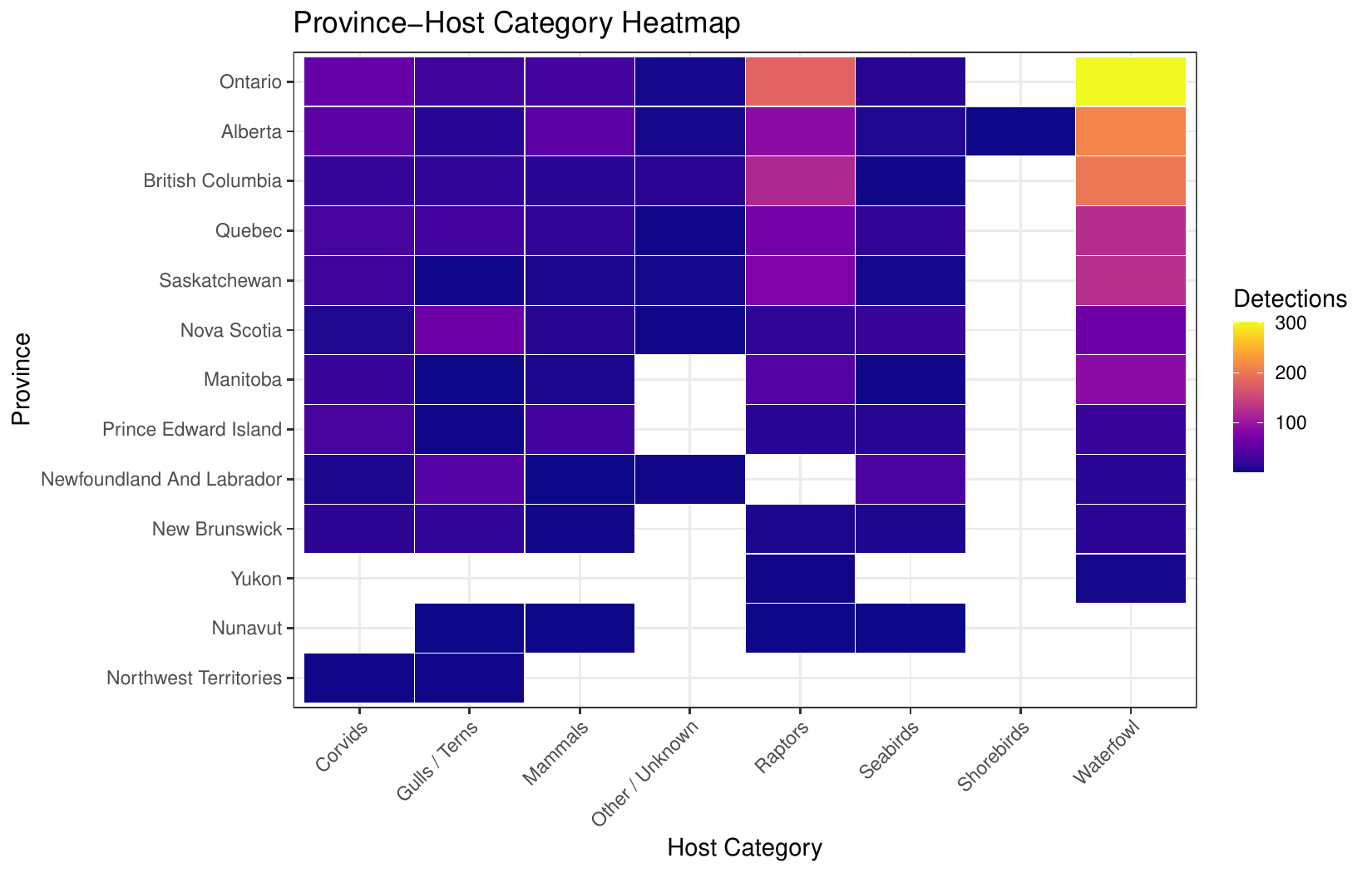}
	\caption{Province-host category heatmap of HPAI A(H5N1) detections in Canadian wildlife. Detection counts were aggregated by province and host category, showing geographic variation in host-category contributions.}
	\label{fig:supp_province_host_category_heatmap}
\end{figure}
\FloatBarrier


\clearpage
\begin{figure}[H]
	\centering
	\includegraphics[width=0.90\textwidth]{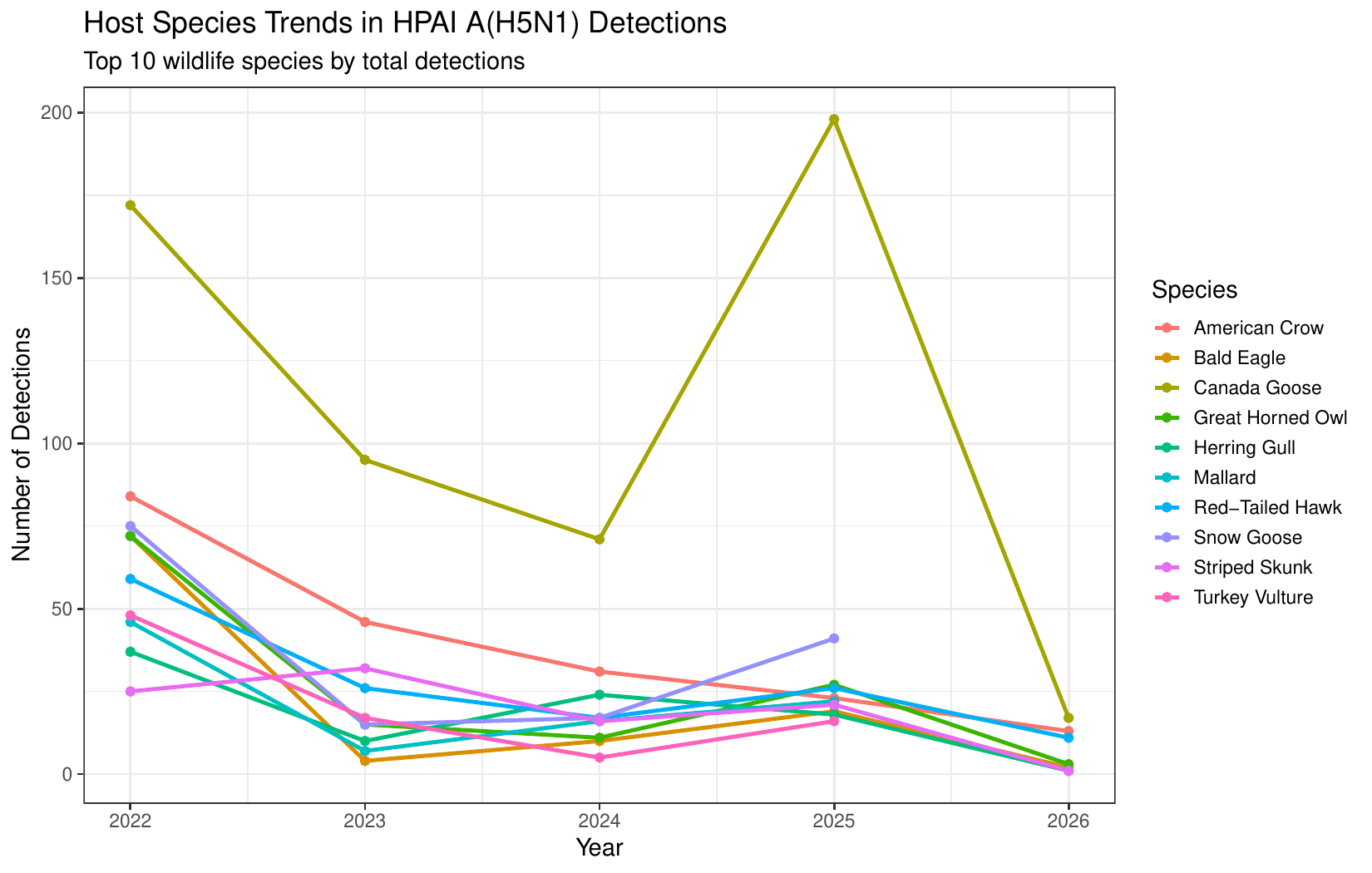}
	\caption{Yearly trends in HPAI A(H5N1) detections among the top 10 wildlife species by total detections. The figure shows species-level variation in detection patterns across surveillance years.}
	\label{fig:supp_host_species_yearly_trends}
\end{figure}
\FloatBarrier

\clearpage

\begin{figure}[H]
	\centering
	\includegraphics[width=0.95\textwidth]{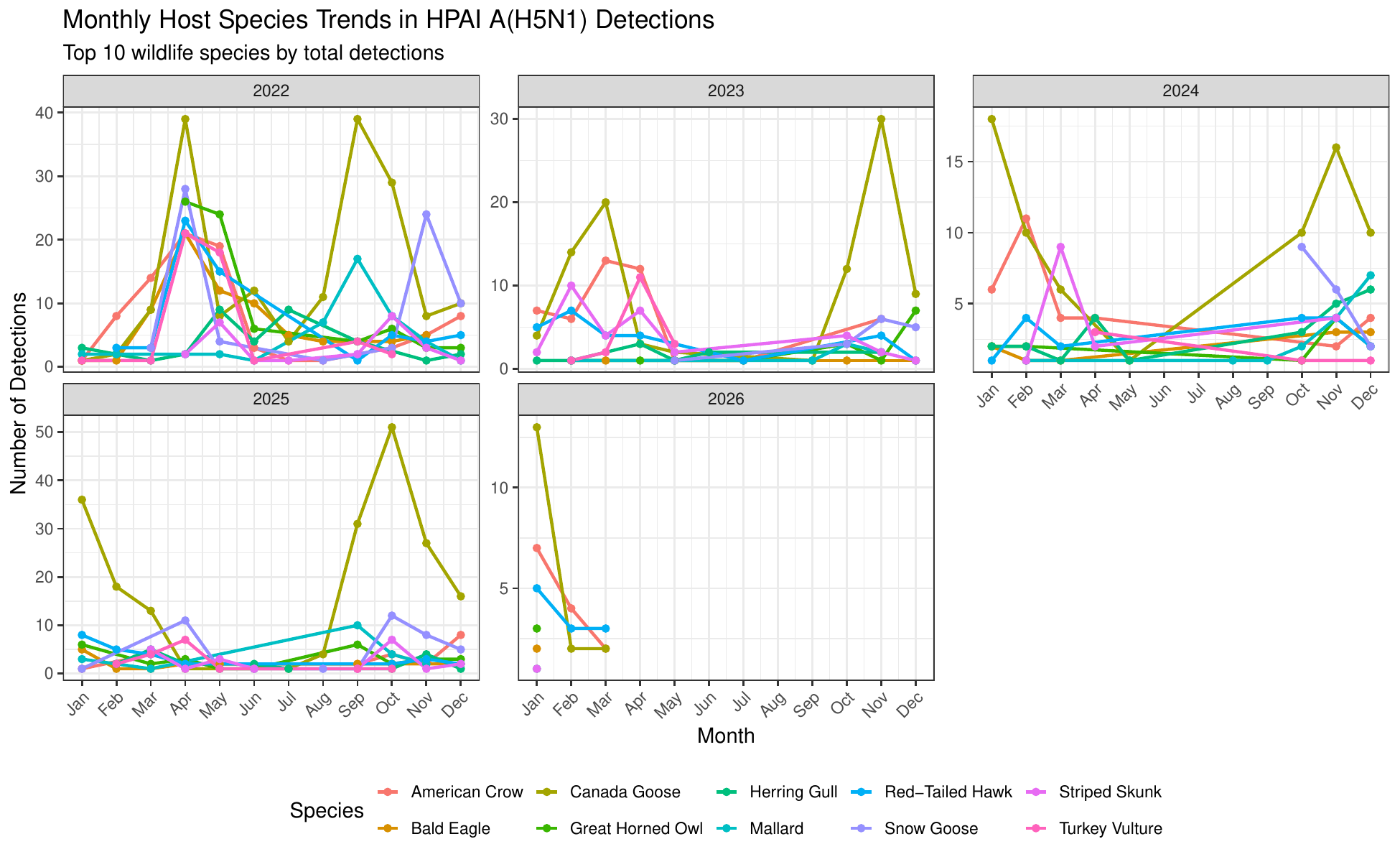}
	\caption{Monthly trends in HPAI A(H5N1) detections among the top 10 wildlife species by total detections, faceted by surveillance year. The figure shows seasonal variation in species-specific detection patterns.}
	\label{fig:supp_host_species_monthly_trends}
\end{figure}
\FloatBarrier


\clearpage
\begin{figure}[H]
	\centering
	\includegraphics[width=0.95\textwidth]{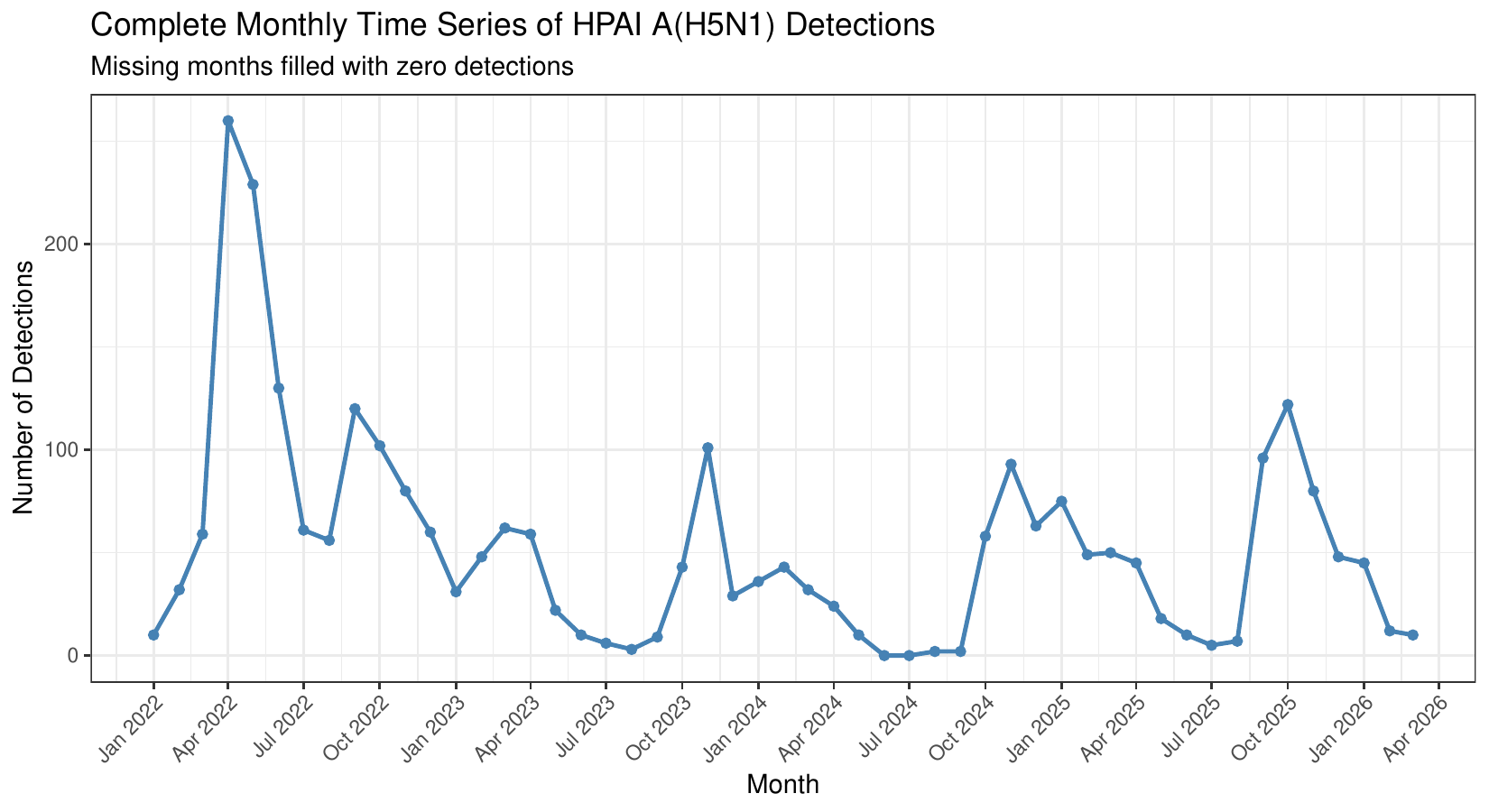}
	\caption{Complete monthly time series of HPAI A(H5N1) wildlife detections in Canada. Monthly detection counts were aggregated across the study period, with missing months filled as zero detections to support time-series analysis.}
	\label{fig:supp_complete_monthly_time_series}
\end{figure}
\FloatBarrier


\clearpage
\begin{figure}[H]
	\centering
	\includegraphics[width=0.95\textwidth]{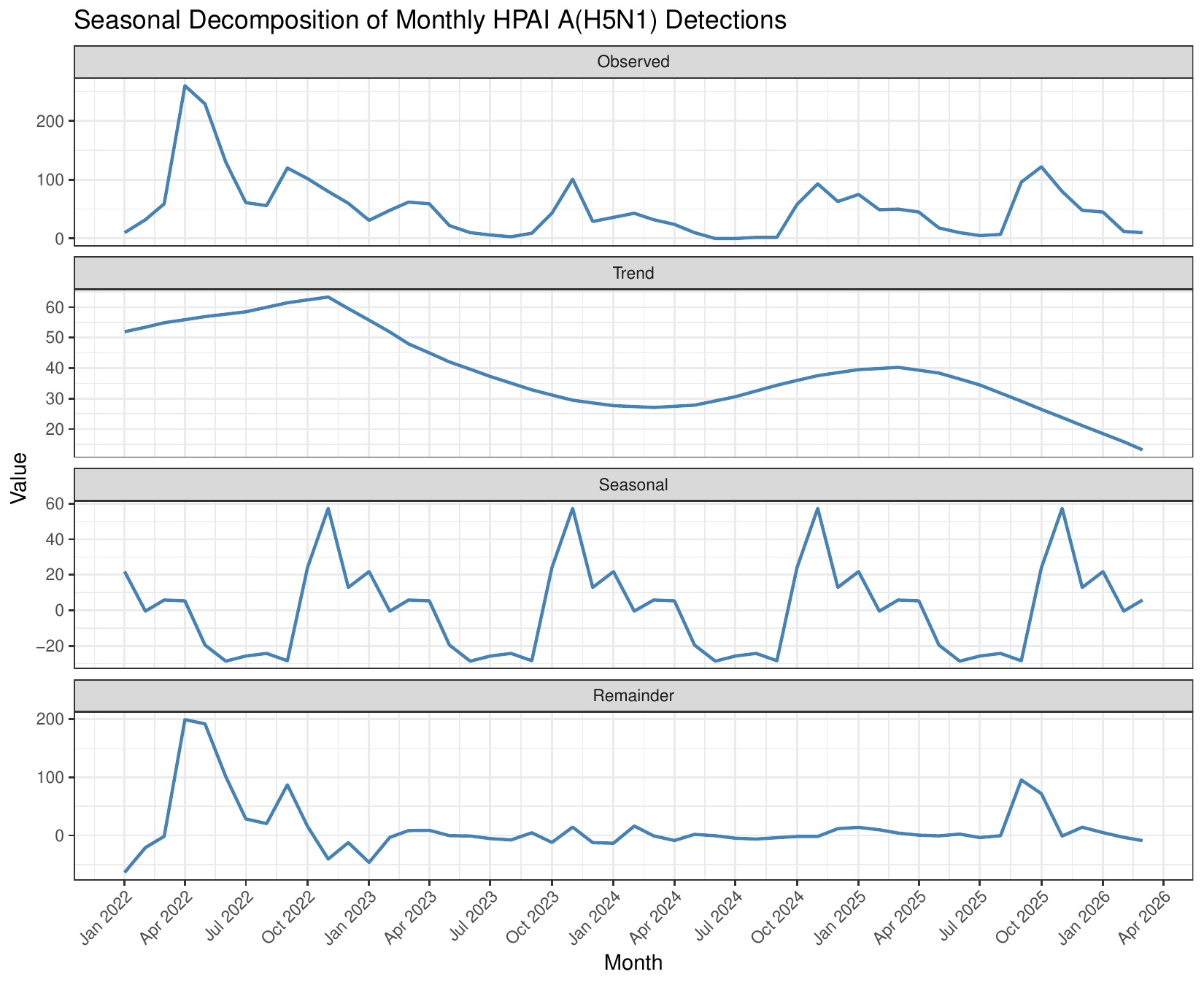}
	\caption{Seasonal decomposition of monthly HPAI A(H5N1) detections in Canadian wildlife. The decomposition separates observed monthly detection counts into trend, seasonal, and remainder components.}
	\label{fig:supp_seasonal_decomposition}
\end{figure}
\FloatBarrier


\clearpage
\begin{figure}[H]
	\centering
	\includegraphics[width=0.85\textwidth]{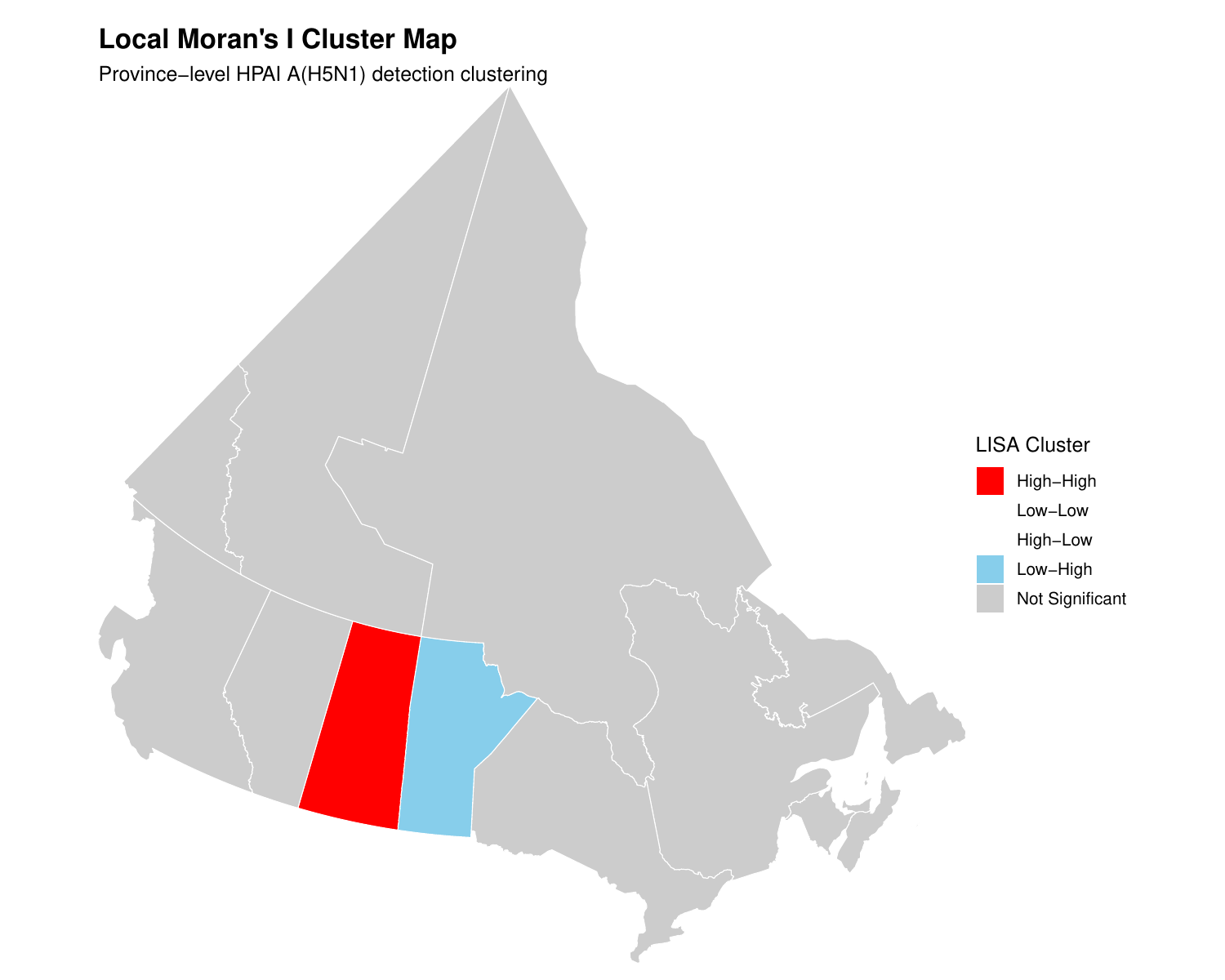}
	\caption{Local Moran's I cluster map of province-level HPAI A(H5N1) detection counts in Canadian wildlife. Provinces and territories were classified as high-high, low-low, high-low, low-high, or not significant based on local spatial association. Results should be interpreted as exploratory because province-level clustering may be influenced by surveillance intensity, province size, wildlife density, and reporting differences.}
	\label{fig:supp_lisa_cluster_map}
\end{figure}
\FloatBarrier


\clearpage
\section*{Supplementary Model Diagnostic Figures}


\begin{figure}[H]
	\centering
	\includegraphics[width=0.90\textwidth]{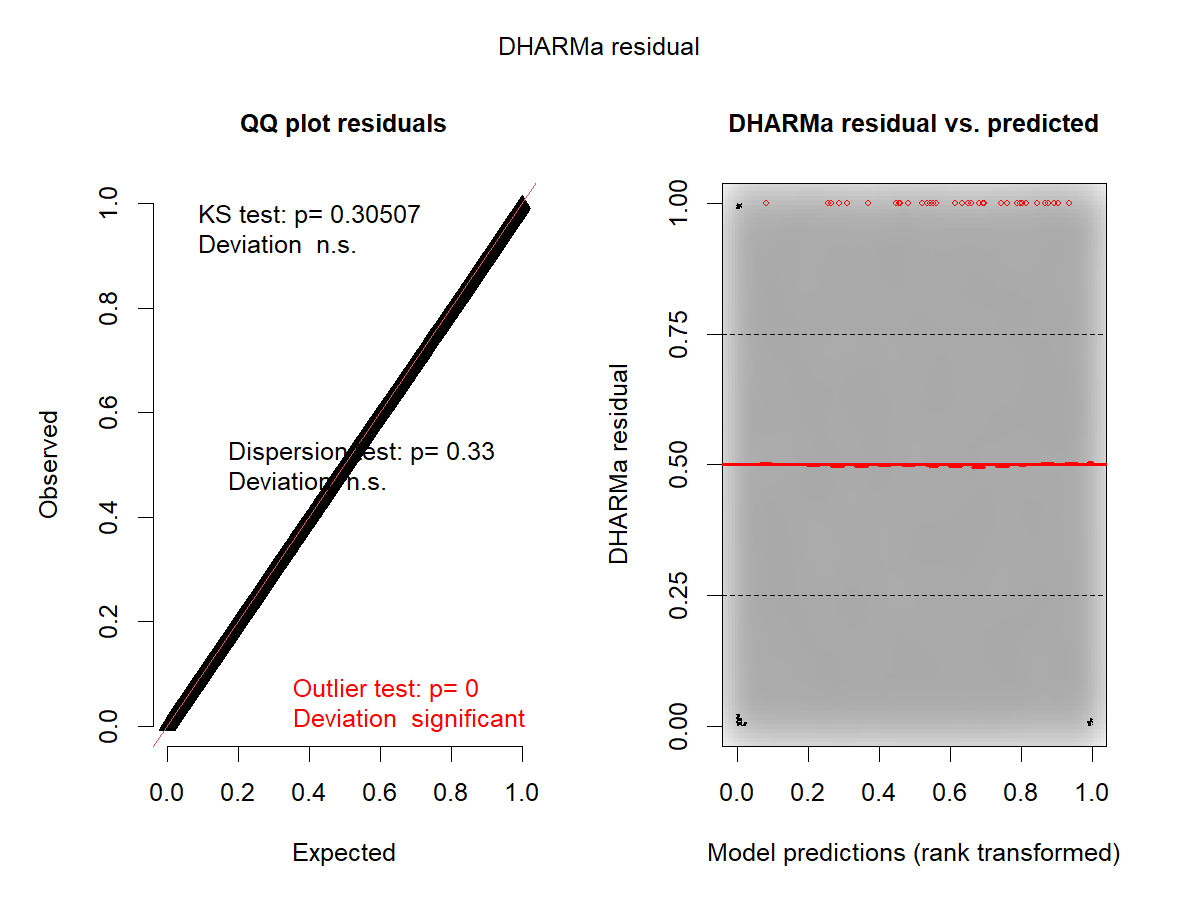}
	\caption{Simulation-based residual diagnostic plots for the final Negative Binomial generalised linear mixed model. Diagnostics were generated using simulated residuals. The model showed no evidence of problematic overdispersion and no evidence of residual zero inflation.}
	\label{fig:supp_diagnostic_nb_season}
\end{figure}
\FloatBarrier


\clearpage

\begin{figure}[H]
	\centering
	
	\begin{subfigure}[t]{0.48\textwidth}
		\centering
		\includegraphics[width=\textwidth]{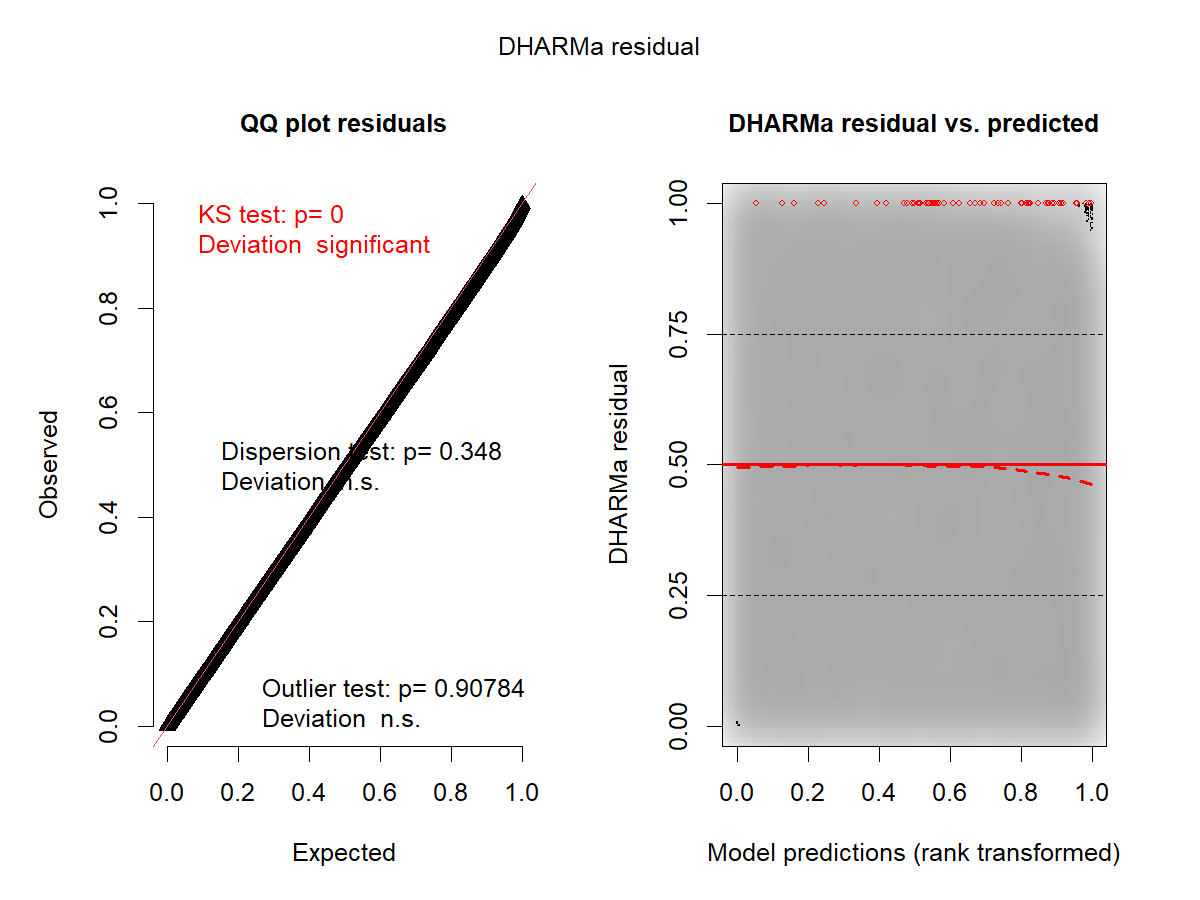}
		\caption{Poisson season model}
		\label{fig:supp_diag_poisson}
	\end{subfigure}
	\hfill
	\begin{subfigure}[t]{0.48\textwidth}
		\centering
		\includegraphics[width=\textwidth]{diagnostic_nb_season.png}
		\caption{Negative Binomial season model}
		\label{fig:supp_diag_nb}
	\end{subfigure}
	
	\vspace{0.4cm}
	
	\begin{subfigure}[t]{0.48\textwidth}
		\centering
		\includegraphics[width=\textwidth]{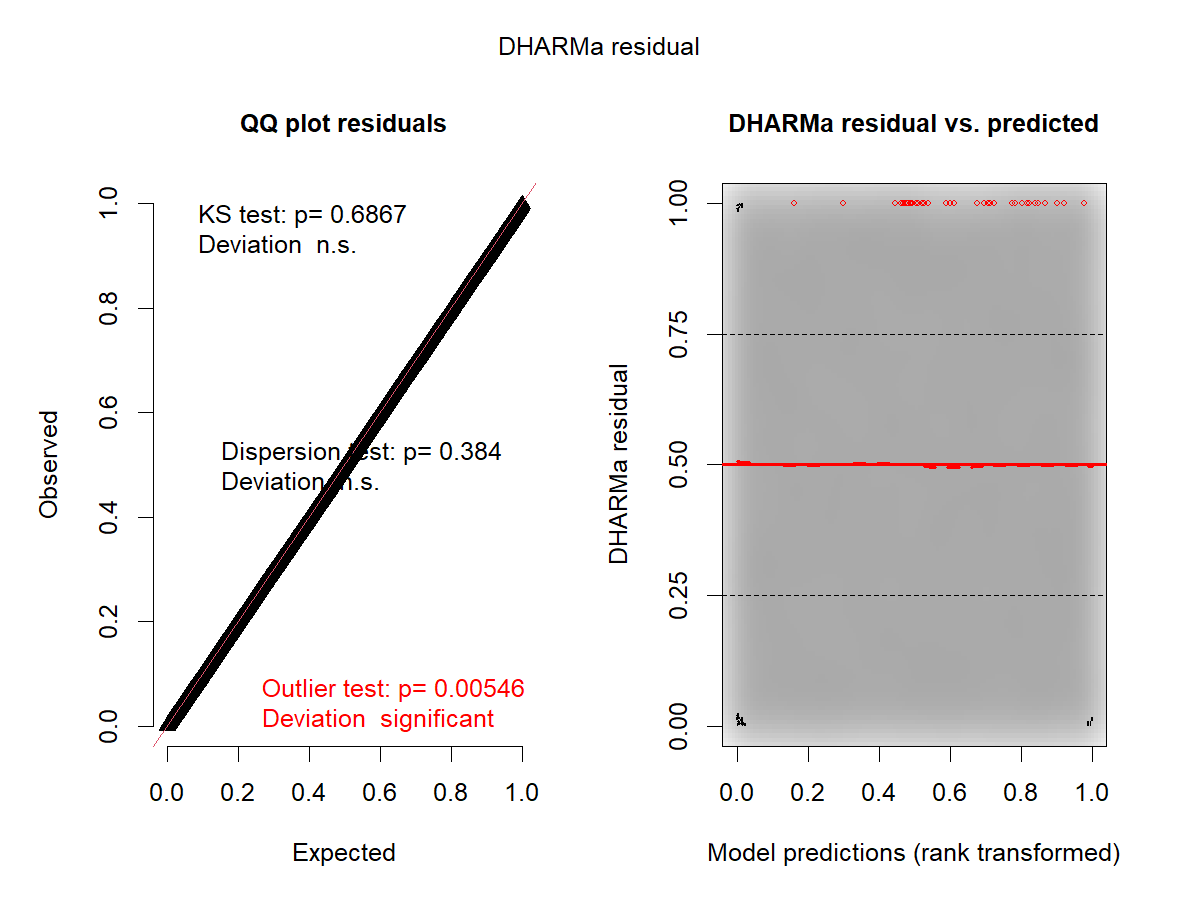}
		\caption{Negative Binomial migration model}
		\label{fig:supp_diag_nb_migration}
	\end{subfigure}
	\hfill
	\begin{subfigure}[t]{0.48\textwidth}
		\centering
		\includegraphics[width=\textwidth]{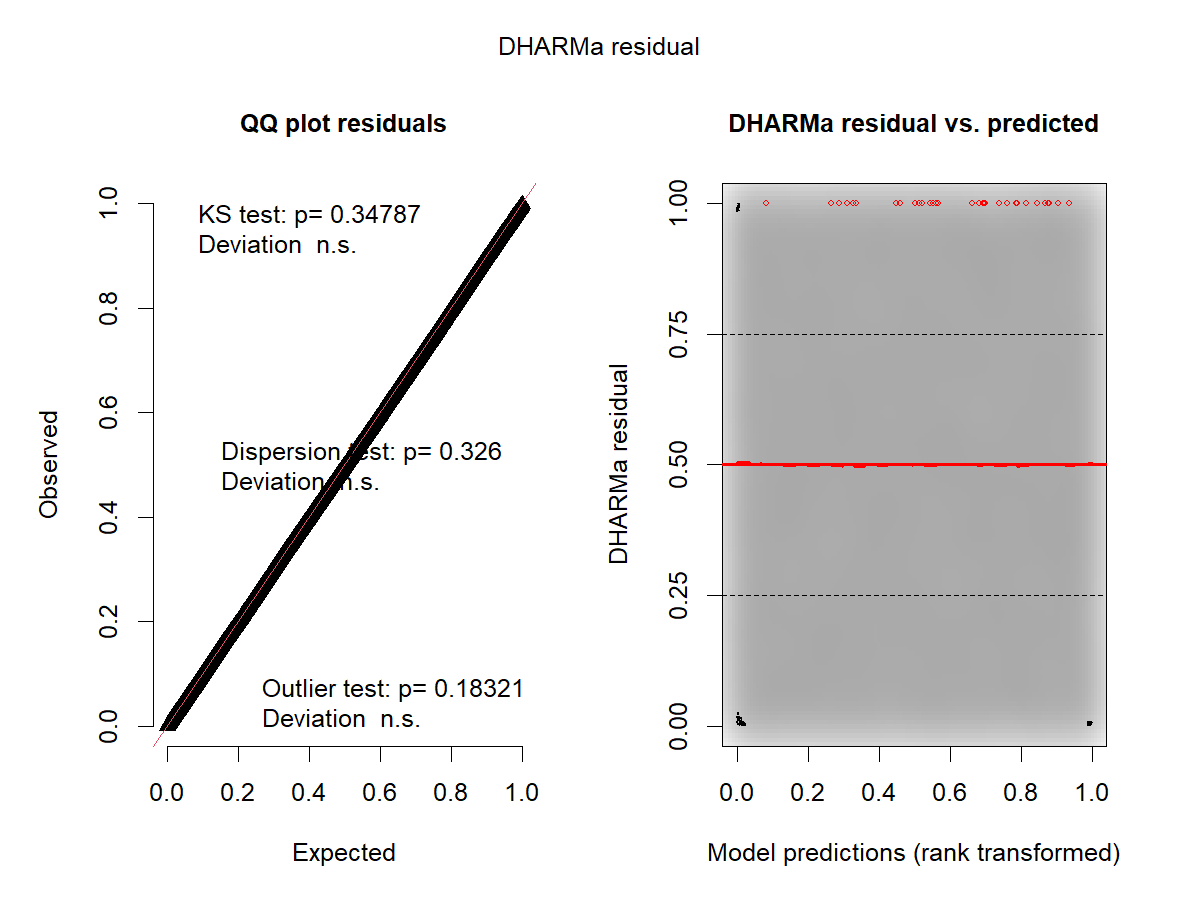}
		\caption{Zero-inflated Negative Binomial season model}
		\label{fig:supp_diag_zinb}
	\end{subfigure}
	
	\caption{Simulation-based residual diagnostic plots for candidate HPAI A(H5N1) detection-count models. Panels show diagnostics for the Poisson season model, Negative Binomial season model, Negative Binomial migration model, and zero-inflated Negative Binomial season model. The Negative Binomial season model showed no evidence of problematic overdispersion or residual zero inflation and was retained as the primary model.}
	\label{fig:supp_model_diagnostics_panel}
\end{figure}
\FloatBarrier


\clearpage
\section*{Supplementary Notes}

\subsection*{Supplementary Note S1. Interpretation of Detection Counts}

The analyses in this study are based on surveillance detections rather than true infection prevalence. Detection counts may be influenced by surveillance intensity, carcass detection probability, public reporting, wildlife population density, laboratory submission practices, and provincial differences in monitoring systems. Therefore, spatial and temporal patterns should be interpreted as detection patterns rather than direct estimates of infection risk.

\subsection*{Supplementary Note S2. Interpretation of Zero-Filled Modelling Dataset}

The risk-factor modelling dataset was constructed as a zero-filled analytical grid across province, year, season, migration period, host category, species group, host type, and lineage. This structure allowed the Negative Binomial model to account for zero-count strata and overdispersion in detection counts. However, some zero-count strata may represent unsampled or under-sampled combinations rather than confirmed absence of infection.

\subsection*{Supplementary Note S3. Host-Type and Host-Category Redundancy}

Host type and host category partially overlapped because the derived host-category variable included a mammal category. As a result, the mammal host-type term was not estimable in the final model. Future models could avoid this redundancy by using either host type or host category, or by adopting a simplified host classification.

\subsection*{Supplementary Note S4. Interpretation of Lineage Effects}

Lineage-associated incidence rate ratios should be interpreted relative to the model reference lineage group. Because some lineage categories were rare or sparsely represented, large IRR estimates may partly reflect sparse counts in less frequent lineage strata rather than only biological differences in transmissibility or detection probability.



\end{document}